\voffset=0.0in
\hoffset=0.25in
\vsize=8.0in
\hsize=6.0in
\def\atsigns#1{\if T#1\catcode `@=11\else \catcode `@=12\fi}
\atsigns{T}
\newdimen\margtitl				
\def\titp@ge{\box\p@gebox\kern-\p@gedp\vskip\botnumbersize} 
\def\nrmp@ge{\botnumberbox\box\p@gebox}                     
\def\titlepage{\gdef\tp@ge{T}}
\def\botnumberbox{\vbox to\botnumbersize{
     \vfil\rightline{\rm \number\pageno .}\vskip20pt}}      
\def\titulo#1{\titlepage
    \topinsert\vskip1.5cm\endinsert {\baselineskip24pt \biggfnt
    \noindent #1}
    \vskip.5cm}                                              
\newbox\h@ngbox					
\def\hangobj#1{\par\setbox\h@ngbox\hbox{#1}\noindent
	       \hangindent=\wd\h@ngbox\box\h@ngbox\hskip\z@\ignorespaces}

\newdimen\footaboveskip				
\newinsert\footins

\def\nfootnote{\global\advance\footno\@ne
	       \footnote{\footnotetag{\number\footno}}}
\def\nvfootnote{\global\advance\footno\@ne
		\vfootnote{\footnotetag{\number\footno}}}

\def\footnote#1{\let\@sf\empty 
		\ifhmode\edef\@sf{\spacefactor\the\spacefactor}\/\fi
		#1\@sf\vfootnote{#1}}
\def\vfootnote#1{\insert\footins\bgroup
		 \interlinepenalty=\interfootnotelinepenalty
		 \splittopskip=\ht\strutbox 
		 \splitmaxdepth=\dp\strutbox \floatingpenalty\@MM
		 \leftskip\z@skip \rightskip\z@skip
		 \spaceskip\z@skip \xspaceskip\z@skip
		 \parskip\z@skip
		 \kern\footaboveskip\footnotefmt{#1}\strut
		 \futurelet\next\fo@t}
\def\fo@t{\ifcat\bgroup\noexpand\next \let\next\f@@t
	  \else\let\next\f@t\fi \next}
\def\f@@t{\bgroup\aftergroup\@foot\let\next}
\def\f@t#1{#1\@foot}
\def\@foot{\strut\egroup}

\skip\footins=15pt		
\count\footins=1000		
\dimen\footins=4in		
\def\footnoterule{\hrule width 1.5in\kern\thr@@\p@}
\footaboveskip=3.5pt		

\def\footnotefam{\eightpoint}
\def\footnotetag#1{{\footnotefam\raise.9ex\hbox{#1}}}
\def\footnotefmt#1{\hangobj{#1\kern\@ne\p@}\footnotefam}
%
%
%
%
%
\atsigns{T}
%
\font\tenrm=cmr10
\font\eightrm=cmr8
\font\sixrm=cmr6
\font\fiverm=cmr5

\font\teni=cmmi10       \skewchar\teni='177
\font\eighti=cmmi8      \skewchar\eighti='177
\font\sixi=cmmi6        \skewchar\sixi='177
\font\fivei=cmmi5       \skewchar\fivei='177

\font\tensy=cmsy10      \skewchar\tensy='60
\font\eightsy=cmsy8     \skewchar\eightsy='60
\font\sixsy=cmsy6       \skewchar\sixsy='60
\font\fivesy=cmsy5      \skewchar\fivesy='60

\font\tensl=cmsl10
\font\eightsl=cmsl8

\font\tenbf=cmbx10
\font\eightbf=cmbx8
\font\sixbf=cmbx6
\font\fivebf=cmbx5

\font\tentt=cmtt10      \hyphenchar\tentt=-1
\font\eighttt=cmtt8     \hyphenchar\eighttt=-1

\font\tenit=cmti10
\font\eightit=cmti8

\font\tensc=cmcsc10
\font\eightsc=cmcsc10   

\global\font\twelverm=cmr10 scaled 1200
\global\font\twelvei=cmmi10 scaled 1200         \global\skewchar\twelvei='177
\global\font\twelvesy=cmsy10 scaled 1200        \global\skewchar\twelvesy='60
\global\font\twelvesl=cmsl10 scaled 1200
\global\font\twelvebf=cmbx10 scaled 1200
\global\font\twelvett=cmtt10 scaled 1200        \global\hyphenchar\twelvett=-1
\global\font\twelveit=cmti10 scaled 1200
\global\font\twelvesc=cmcsc10 scaled 1200
%
%
%
%
\newfam\capsfam

\gdef\twelvepoint{
	\def\rm{\fam\z@\twelverm}\textfont\z@=\twelverm
				 \scriptfont\z@=\tenrm
				 \scriptscriptfont\z@=\eightrm
	\def\it{\fam\itfam\twelveit}\textfont\itfam=\twelveit
	\def\sl{\fam\slfam\twelvesl}\textfont\slfam=\twelvesl
	\def\bf{\fam\bffam\twelvebf}\textfont\bffam=\twelvebf
				    \scriptfont\bffam=\tenbf
				    \scriptscriptfont\bffam=\eightbf
	\def\tt{\fam\ttfam\twelvett}\textfont\ttfam=\twelvett
        \def\caps{\fam\capsfam\twelvesc}\textfont\capsfam=\twelvesc
	\textfont\@ne=\twelvei	\scriptfont\@ne=\teni
				\scriptscriptfont\@ne=\eighti
	\textfont\tw@=\twelvesy \scriptfont\tw@=\tensy
				\scriptscriptfont\tw@=\eightsy
	\setbox\strutbox=\hbox{\vrule height10.2\p@ depth4.2\p@ width\z@}%
	\normalbaselineskip=14.4\p@\normalbaselines\rm}%
\def\eightpoint{
	\def\rm{\fam\z@\eightrm}\textfont\z@=\eightrm
				\scriptfont\z@=\sixrm
				\scriptscriptfont\z@=\fiverm
	\def\it{\fam\itfam\eightit}\textfont\itfam=\eightit
	\def\sl{\fam\slfam\eightsl}\textfont\slfam=\eightsl
	\def\bf{\fam\bffam\eightbf}\textfont\bffam=\eightbf
				   \scriptfont\bffam=\sixbf
				   \scriptscriptfont\bffam=\fivebf
	\def\tt{\fam\ttfam\eighttt}\textfont\ttfam=\eighttt
	\def\caps{\fam\capsfam\eightsc}\textfont\capsfam=\eightsc
	\textfont\@ne=\eighti  \scriptfont\@ne=\sixi
			       \scriptscriptfont\@ne=\fivei
	\textfont\tw@=\eightsy \scriptfont\tw@=\sixsy
			       \scriptscriptfont\tw@=\fivesy
	\setbox\strutbox=\hbox{\vrule height7\p@ depth\tw@\p@ width\z@}%
	\normalbaselineskip=9\p@\normalbaselines\rm}
\def\tenpoint{
	\def\rm{\fam\z@\tenrm}\textfont\z@=\tenrm
			      \scriptfont\z@=\eightrm
			      \scriptscriptfont\z@=\sixrm
	\def\it{\fam\itfam\tenit}\textfont\itfam=\tenit
	\def\sl{\fam\slfam\tensl}\textfont\slfam=\tensl
	\def\bf{\fam\bffam\tenbf}\textfont\bffam=\tenbf
				 \scriptfont\bffam=\eightbf
				 \scriptscriptfont\bffam=\sixbf
	\def\tt{\fam\ttfam\tentt}\textfont\ttfam=\tentt
	\def\caps{\fam\capsfam\tensc}\textfont\capsfam=\tensc
	\textfont\@ne=\teni  \scriptfont\@ne=\eighti
			     \scriptscriptfont\@ne=\sixi
	\textfont\tw@=\tensy \scriptfont\tw@=\eightsy
			     \scriptscriptfont\tw@=\sixsy
	\setbox\strutbox=\hbox{\vrule height8.5\p@ depth3.5\p@ width\z@}%
	\normalbaselineskip=12\p@\normalbaselines\rm}

%
\atsigns{F}
\twelvepoint

\font\biggfnt=cmr10 scaled \magstep3

\newfam\msbfam
\font\twlmsb=msbm10 at 12pt
\font\eightmsb=msbm10 at 8pt
\font\sixmsb=msbm10 at 6pt
\textfont\msbfam=\twlmsb
\scriptfont\msbfam=\eightmsb
\scriptscriptfont\msbfam=\sixmsb
\def\cj{\fam\msbfam}

\def\C{{\cj C}}

\def\R{{\cj R}}

\def\Z{{\cj Z}}
\def\section#1{\vskip10pt\noindent {\bf \uppercase{#1}}}
\def\subsection#1{\vskip7.5pt\noindent {\bf #1}}
\def\aderecha#1{\smash{\mathop{\longrightarrow}\limits^{#1}}}


\def\ro{\mathop{\tilde{\rho}}\limits^{\circ}}

\vglue2.5truecm

\titulo{The Universal Covering Group of U(n) and Projective Representations}

\bigskip

\hangindent=1.5cm
{\bf M.$\!$ A. Aguilar\footnote{$^{1}$}{Instituto de
Matem\'aticas, Universidad Nacional Aut\'onoma de M\'exico, Circuito
Exterior, Cd.  Universitaria, 04510 M\'exico, D.$\!$~F., M\'exico; e-mail:
marcelo@math.unam.mx} and M. Socolovsky\footnote{$^{2}$}{Instituto de
Ciencias Nucleares, Universidad Nacional Aut\'onoma de M\'exico,
Circuito Exterior, Cd. Universitaria, 04510 M\'exico,
D.$\!$~F., M\'exico; e-mail: socolovs@nuclecu.unam.mx}}

\bigskip

\bigskip

\hangindent=1.5cm {\tenpoint
\hskip.8cm
Using fibre bundle theory we construct the universal covering group of  
$U(n)$, $\tilde{U}(n)$, and show that $\tilde{U}(n)$ is isomorphic to 
the semidirect product $SU(n)\bigcirc \hskip-8pt{\scriptstyle s}$ $\R$. 
We give a bijection between the set of projective representations of $U(n)$ 
and the set of equivalence classes of certain unitary representations of 
$SU(n)\bigcirc \hskip-10pt{\scriptstyle s}$ $\R$. 
Applying Bargmann's theorem, 
we give explicit expressions for the liftings of projective representations 
of $U(n)$ to unitary representations of $SU(n)\bigcirc \hskip-10pt{\scriptstyle s}$ 
$\R$. For completeness, we discuss the topological and group theoretical relations between $U(n)$, $SU(n)$, 
$U(1)$ and $\Z _n$.} 

\bigskip

\section{1. Introduction}

\

The universal covering  group $\tilde{G}$ of a connected Lie group $G$ (Warner, 1983), 
and its associated projection $\tilde{G} \to G$, play in many cases an important 
r\^ole in physical applications. This is due to the fact that $\tilde{G}$ has 
representations which do not come from representations of $G$, the so called 
spinor representations of $G$. The most important and well known examples are 
the rotations in 3-dimensional Euclidean space, $SU(2)\to SO(3)$ and the Lorentz 
transformations in 4-dimensional Minkowski spacetime, $SL_2(\C)\to SO^0(3,1)$. 

In the abelian case, the universal covering group of the circle, 
$\R \to U(1)$ is the vehicle through which virtual classical paths contribute to the 
Feynman path integral expression for the quantum transition amplitudes in non-
relativistic quantum mechanics (Feynman and Hibbs, 1965). 

On the other hand, projective representations of symmetry groups appear naturally in 
quantum mechanics, since, on the one hand, the pure states of any physical system are 
represented by {\it rays} in the corresponding Hilbert space, and, on the other, 
the unitary or antiunitary operators representing the symmetry transformations are 
required to preserve transition probabilities, that is, the square of the modulus of the 
transition amplitudes, but not the transition amplitudes themselves; therefore the operators are   determined only up to a phase. For Lie groups, however, these extra phases can be eliminated, that is, one can pass to a truly unitary representation, when the symmetry group satisfies the conditions of Bargmann's theorem (Bargmann, 1954), namely, the group should be simply connected and the 2nd cohomology of its Lie algebra should be equal to zero (see Section 4). If the group is connected but not simply connected, one passes to its universal covering group.

One should emphasize, however, that the quantum physical symmetry group is determined by the projective representation (Belinfante, 1972).

In this paper we study the universal covering group $\tilde{U}(n)$ of $U(n)$, the automorphism group 
of the Hilbert space $\C^n$, and show that $\tilde{U}(n)$ is isomorphic to the 
semidirect product $SU(n)\bigcirc \hskip-11pt{\scriptstyle s}$ $\R$ (Section 3). From the physical point of view, $U(n)$ is the internal symmetry group of a system of $n$ identical noninteracting  harmonic oscillators (Bargmann and Moshinsky, 1960). 

In Section 4 we discuss some preliminary results on unitary and projective representations. 
In particular, we study the concept of strong continuity of a representation, and show that 
the group of unitary transformations of a Hilbert space $\cal H$, $\cal U (\cal H)$, is a topological group. 

In Section 5 we give a bijection between the set of projective representations of $U(n)$ and the set of orbits of certain unitary representations of $SU(n)\bigcirc \hskip-11pt{\scriptstyle s}$ $\R$ under the action of the group of one-dimensional representations of $SU(n) \bigcirc \hskip-11pt{\scriptstyle s}$ $\R$. 

In Section 6 we construct all the unitary representations of $\tilde{U}(n)$ associated to a projective representation of $U(n)$. 

As an introduction, we discuss, 
in Section 2, the topological and group theoretical relations between $U(n)$, $U(1)$, 
$SU(n)$ and $\Z_n$. 

\section{2. The groups $U(n)$} 

\

The unitary group $U(n)$, $n=$1, 2, ..., is the group of automorphisms of 
the Hilbert space $(\C^n,<$ , $>)$ where $< $, $>$ is the Hermitian scalar  
product $<\vec{z},\vec{w}>=\sum\limits^n_{k=1}z_k \bar w_k$. If $A\in U(n)$ 
and $A^*$ is the transpose conjugate matrix, then $A^*A=I$ {\it i.e.} $A^*
=A^{-1}$, so $\vert \det A \vert =1$ and $\dim_{\R}U(n)=n^2$. $U(n)$ is a 
Lie group and $SU(n)$ is the closed Lie subgroup consisting of matrices 
whose determinant is 1. In particular $U(1)$ is the unit circle or 1-sphere 
$S^1$.

Let $\varphi_n:U(1) \times SU(n) \to U(n)$ be given by $\varphi_n(z,A):=zA.$ 
Since $(zA)^*= \bar z A^*=(zA)^{-1}$, $\varphi_n$ is well defined. 
Now we show that $\varphi_n$ 
is a surjective Lie group homomorphism with kernel $\Z_n$: i)  
$\varphi_n ((z,A)(z^\prime,A^\prime))=\varphi_n(zz^\prime,AA^\prime)=
(zz^\prime)(AA^\prime)=(zA)(z^\prime A^\prime)=\varphi_n(z,A)\varphi_n
(z^\prime,A^\prime)$ {\it i.e.} $\varphi_n$ is a group homomorphism; 
ii) if $B\in U(n)$ then $(detB)^{-1/n}B\in SU(n)$ and 
$\varphi_n ((detB)^{1/n}, (detB)^{-1/n}B)=B$ {\it i.e.} $\varphi_n$ is onto; 
iii) let $(e^{-i\theta}, (
a_{ij}))$ be in $ker(\varphi_n)$ {\it i.e.} $e^{-i\theta}(a_{ij})=I$, then 
$a_{ij}=0$ for $i\neq j$ and $a_{11}=$...$=a_{nn}=e^{i\theta}$; since 
$det(a_{ij})=e^{in\theta}=1$ then $\theta=2\pi m/n$ for $m=$ 0, 1, ..., 
$n-1$ and therefore $ker (\varphi_n)=\{(1,I), (e^{-i2\pi/n}, 
e^{i2\pi/n} I),$ ..., $(e^{-i2\pi(n-1)/n},e^{i2\pi(n-1)/n} I)\}\cong \Z_n$.  
Then one has an isomorphism of short exact sequences of Lie groups and Lie group 
homomorphisms given by the following diagram:
$$\matrix{1 & \to & \Z _n & \aderecha{\iota_n} & U(1)\times SU(n) &
\aderecha{\varphi_n} & U(n) &\to & 1 \cr
& & \| & & \| & & \Phi_n\uparrow \cong & & \cr
1 & \to & \Z_n & \aderecha{\iota_n} & U(1)\times SU(n) & \aderecha{p_n}
& {U(1)\times SU(n)}\over {\Z _n} & \to & 1 \cr}$$
where $\iota_n$ is the inclusion $\iota_n(k)=(e^{-i2\pi k/n}, 
e^{i2\pi k/n}I)$, $k=$0, 1, ..., $n-1$, $p_n$ is the canonical projection 
$p_n(z,A)= [z,A]$, and $\Phi_n$ is the {\it 
Lie group isomorphism} $\Phi_n ([z,A])=\varphi_n (z,A)=zA$.

On the other hand, $d:U(n) \to U(1)$ given by $A \mapsto d(A):= det A$ is 
a Lie group homomorphism with $ker(d)=SU(n)$. Then one has another isomorphism 
of short exact sequences of Lie groups and Lie group homomorphisms given by the
following diagram:
$$\matrix{1 & \to & SU(n) &\aderecha{\iota} & U(n) & \aderecha{d} & U(1) & 
\to & 1 \cr & & \| & & \| & & \cong \downarrow \psi & & \cr
1 & \to & SU(n) & \aderecha{\iota} & U(n) & \aderecha{\pi} & {U(n)}\over {SU(n)} 
& \to & 1 \cr}$$
where $\iota$ is the inclusion $\iota(A)=A$, $\pi $ is the canonical 
projection $\pi(B)=BSU(n)$, and $\psi $ is the group isomorphism $\psi (z)=
\pmatrix{z & 0 \cr 0 & I \cr}SU(n)$ where $I$ is the $(n-1)\times 
(n-1)$ unit matrix. Since $\pi$ is a principal $SU(n)-$ bundle, 
then, $U(n) \aderecha{d} U(1)$ is an $SU(n)$-principal 
bundle over $S^1$, and since the set of isomorphism classes of principal 
$SU(n)$-bundles over the 1-sphere, $k_{SU(n)}(S^1)$ is in one-to-one 
correspondence with $\Pi_0(SU(n)) \cong 0$, then the bundle $d$ is trivial. 
The global section $\sigma :U(1) \to U(n)$, $z \mapsto \sigma (z) := 
\pmatrix{z & 0 \cr 0 & I \cr}$ induces the $SU(n)$-{\it bundle 
isomorphism}
$$\matrix{SU(n) & & SU(n) \cr
\downarrow & & \downarrow \cr
U(n) & \aderecha{\Psi_n} & U(1)\times SU(n) \cr
d\searrow & & \swarrow \pi_1 \cr
& U(1) & \cr}$$
given by $\Psi_n (B)=(z,A)$ with $z=det B$ and $A=\sigma(\bar z)B$; where 
$\pi_1(z,A)=z$. $\Psi_n$ is {\it not} a Lie group homomorphism ($
\Psi_n(BB^\prime)=(zz^\prime, \sigma(\bar z)\sigma(\bar {z^\prime})
BB^\prime)\neq \Psi_n (B) \Psi_n(B^\prime)$ since for all $B \in U(n)$, 
$\sigma (\bar {z^\prime})B = B\sigma (\bar {z^\prime})$ only if $z^\prime =1$) 
but only a {\it diffeomorphism} of smooth manifolds; its inverse is 
given by $\Psi_n^{-1}(z,A)=\sigma (z)A$.

\

In summary, we have the commutative diagram
$$\matrix{& {U(1)\times SU(n)}\over {\Z _n}  & \cr
\Phi_n \swarrow  & & \searrow t_n \cr
U(n) & \aderecha{\Psi_n} & U(1)\times SU(n) \cr}$$
where $t_n=\Psi_n \circ \Phi_n$ is given by $t_n([z,A])=(z^n, \sigma ( 
\bar z ^n)zA)$ and is a {\it diffeomorphism} of smooth manifolds but {\it 
not} a Lie group isomorphism. So, $\varphi_n :U(1)\times SU(n) \to U(n)$ 
is an $n$-covering space of $U(n)$ by 
a space diffeomorphic to it. (This result is similar to that of the double 
covering of the circle, the $\Z _2$-bundle $S^1 \to S^1$, $z \mapsto 
z^2$.)

\section{3. The universal covering group $\tilde{U}(n)$} 

\

Since topologically $U(n)\cong U(1)\times SU(n)$, then $\Pi_1 (U(n)) \cong 
\Pi_1(U(1)\times SU(n))\cong \Pi_1(U(1)) \oplus \Pi_1(SU(n)) \cong \Z 
\oplus 0 \cong \Z$, and so the universal covering group of $U(n)$, 
$\tilde{U}(n)$ is a principal $\Z $-bundle $\xi:\Z \to \tilde{U}(n) \to U(n)$. 
On the other hand, the universal principal $\Z$-bundle is $\xi _\Z :\Z \to 
\R \aderecha{exp} U(1)$ with $exp(t)=e^{i2\pi t}$. Then $\xi$ is a pull-back 
of $\xi _\Z $. We have found a natural map between $U(n)$ and $U(1)$, 
namely the determinant $d$. We shall prove that $$ \tilde{U}(n) \cong 
d^*(\R) \cong SU(n) \bigcirc \hskip-11pt{\scriptstyle s} \hskip10pt{\R}$$

{\it Remark 1.} The first isomorphism has been given by Fulton and Harris 
(Fulton and Harris, 1991), but not its relation to fibre bundle theory. 

\

{\it Proposition 1.} Let $d^*(\xi _\Z)$ be the pull-back bundle of $\xi _
\Z $ by the determinant $d: U(n) \to U(1)$ :
$$\matrix{\Z & & \Z \cr
\downarrow & & \downarrow \cr
d^*(\R) & \aderecha{\pi_2} & \R \cr
\pi_1 \downarrow & & \downarrow exp \cr
U(n) & \aderecha{d} & U(1) \cr}$$
where $\pi_1$ and $\pi_2$ are the projections in the first and second 
factor respectively. Then $\pi_1$ is the universal covering group of $U(n)$.

\

{\it Proof.} Since $d$ and $exp$ are group homomorphisms, $d^*(\R)$ is a 
subgroup of $U(n) \times \R $, and clearly $\pi_1$ and $\pi_2$ are homomorphisms. 
The subgroup $d^*(\R)$ is the subset of $U(n)\times \R$ where the maps $d$ and $exp$ coincide, and $U(1)$ is Hausdorff, threfore $d^*(\R)$ is closed. Since $U(n)\times \R$ is a Lie group, $d^*(\R)$ is also a Lie group. Clearly, the maps $\pi_1$ and $\pi_2$ are smooth. 

Now $exp:\R \to U(1)$ is a covering space with fiber $\Z$. Therefore $\pi_1 :d^*(\R) \to U(n)$ is also a covering space with fiber $\Z$. Hence, we have a monomorphism $(\pi_1)_*:\Pi_1(d^*(\R))\to \Pi_1(U(n))$, and the quotient $\Pi_1(U(n))/(\pi_1)_*(\Pi_1(d^*(\R)))$ is isomorphic to $\Z$. Since $\Pi_1(U(n))\cong \Z$, then $\Pi_1(d^*(\R))\cong 0$.    QED

\

{\it Definition 1.} Consider the Lie groups $SU(n)$ and $\R$. We define a smooth action $SU(n)\times 
\R \aderecha{\cdot} SU(n)$ given by $(A,t)\mapsto A\cdot t:=\sigma_{-t} A\sigma_t$, where $\sigma_t:=\sigma(e^{i2\pi t})$. Using this action we have the semidirect product $SU(n) \bigcirc \hskip-11pt{\scriptstyle s}$ $\R$, whose underlying manifold is $SU(n) \times \R$, but with product given by the formula: $(A,t)(A^\prime,t^\prime):=((A\cdot t^\prime)A^\prime,t+t^\prime)$. One can easily check that this is a Lie group. 

\

{\it Theorem 1.} The universal covering group of $U(n)$ is given by the map $p:SU(n) \bigcirc \hskip-11pt{\scriptstyle s}$ $\R \to U(n)$, where $p(A,t)=\sigma(e^{i2\pi t})A$.

\

{Proof.} As mentioned above, $SU(n) \bigcirc \hskip-11pt{\scriptstyle s}$ $\R$ is a Lie group and a simple calculation shows that $p$ is a homomorphism. Since $p$ is a composition of smooth maps, it is smooth. 

Let $\omega:SU(n) \bigcirc \hskip-11pt{\scriptstyle s}$ $\R \to d^*(\R)$ be the map given by $\omega(A,t)=(p(A,t),t)$. Clearly $\omega$ is smooth and one can easily verify that it is a homomorphism of Lie groups, whose inverse is given by $\omega^{-1}(B,t)=(\sigma(e^{-i2\pi t})B,t)$. Then we have the following commutative diagram:
$$\matrix{SU(n)\bigcirc \hskip-11pt{\scriptstyle s}$ $\R & \aderecha{\omega} & d^*(\R) \cr
p \searrow & & \swarrow \pi_1 \cr
& U(n) & \cr}$$
By Proposition 1, $d^*(\R)$ is the universal covering group of $U(n)$, therefore $SU(n) \bigcirc \hskip-11pt{\scriptstyle s}$ $\R$ is also the universal covering group of $U(n)$.    QED

\

{\it Remark 2.} In the literature one sometimes finds that the universal covering group of $U(n)$ is the direct sum of $SU(n)$ and $\R$ (Cornwell, 1984); however, by the theorem above, this is not the case. 

\

{\it Remark 3.} For the particular case $n=2$, $\Phi_2$ in Section 2 says 
that as a group, $U(2)$ is constructed from the {\it unique } three 
spheres which are groups, namely $S^0$, $S^1$ and $S^3$, respectively the 
unit real, complex and quaternionic numbers (Aguilar, Gitler and Prieto, 1998), and one has the commutative diagram
$$\matrix{& {S^1 \times S^3} \over S^0 & \cr
\Phi_2 \swarrow  & & \searrow t_2 \cr
U(2) & \aderecha{\Psi_2} & S^1 \times S^3 \cr}$$
where $\Phi_2$ is a group isomorphism, and $\Psi_2 $ and $t_2$ are 
diffeomorphisms. The explicit formulae for this case are the following: 
$p_2(z,A)\equiv [z,A]=\{(z,A),(-z,-A)\}$, $\Phi_2([z,A])=zA$, and $\Psi_2
\pmatrix{a & b \cr -\bar b e^{i\lambda} & \bar a e^{i\lambda} \cr}= 
(e^{i \lambda}, \pmatrix{ae^{-i\lambda} & be^{-i\lambda} \cr 
-\bar b e^{i\lambda} & \bar a e^{i\lambda} \cr})$ where $\vert a \vert ^2
+ \vert b \vert ^2=1$ and $\lambda \in [0, 2\pi)$. (Another way to show 
the topological equivalence of $U(2)$ and $U(1) \times SU(2)$ is to 
start from the bundle $U(1) \to U(2) \to U(2)/U(1)$ and use the fact that 
$U(2)/U(1) \cong S^3$; since $\Pi_2(U(1))\cong 0$ then the bundle is 
trivial, and so $U(2) \cong U(1)\times S^3$.) 

The product in the universal covering $\tilde{U}(2)\cong SU(2)
\bigcirc \hskip-11pt{\scriptstyle s}$ $\R$ is given by 
$$(\pmatrix{a & b \cr -\bar b & \bar a \cr}, t)(\pmatrix{a^\prime & b^\prime 
\cr -\bar {b^\prime} & \bar {a^\prime} \cr}, t^\prime )=
(\pmatrix{aa^\prime-b\bar {b^\prime}e^{-i2\pi t^\prime} & 
ab^\prime +b\bar {a^\prime}e^{-i2\pi t^\prime} \cr
-\bar a \bar {b^\prime} -\bar b a^\prime e^{i2\pi t^\prime} &
\bar a \bar {a^\prime} - \bar b b^\prime e^{i2\pi t^\prime} \cr}, t+t^\prime).$$

\section{4. Projective and unitary representations}

\

Let $\cal H$ be a complex Hilbert space with the standard norm topology 
$\vert \vert v \vert \vert ^2 =<v,v>$ for any $v\in \cal H$. Let $\hat{\cal H}$ be the 
projective space of its 1-dimensional subspaces with the quotient topology 
relative to the projection $v \mapsto \hat{v}:=\C^*v$, for $v \neq 0$. Let 
$Aut(\hat{\cal H})$ be the group of 
automorphisms of $\hat {\cal H}$, where $T:\hat {\cal H} \to \hat {\cal H}$ is an automorphism if 
$<T(\hat{v_1}), T(\hat{v_2})>=<\hat{v_1}, \hat {v_2}>$,  
where $<\hat {v_1},\hat{v_2}>$ (the transition probability in quantum mechanics) is given 
by $\vert <v_1,v_2>\vert ^2 / \vert \vert v_1 \vert \vert ^2 \vert \vert v_2 
\vert \vert ^2$. Let $\tilde{\cal U}(\cal H)$ be the group of unitary or antiunitary 
transformations of $\cal H$. Let $\pi:\tilde{\cal U}({\cal H}) \to Aut(\hat{\cal H})$ be the projection 
$\pi (A)(\hat{v})=\hat{A}(\hat{v}):=\widehat{Av}$; and $\iota:U(1) \to \tilde{\cal U}({\cal H})$ the inclusion 
$\iota (z)=zId$. Then Wigner's theorem (Wigner, 1931) says that the following 
sequence is exact: $$1\to U(1)\aderecha{\iota} \tilde{\cal U}({\cal H}) \aderecha{\pi} 
Aut(\hat{\cal H})\to 1 $$ {\it i.e.} any probability preserving transformation of the 
projective Hilbert space is the image of a unitary or antiunitary transformation 
of the Hilbert space itself, and moreover, if $\pi (A_1)=\pi (A_2)$ then $A_2=
e^{i\varphi}A_1$ with $\varphi \in [0,2\pi)$ (Simms, 1968). If ${\cal U}(\cal H)$ is 
the subgroup of $\tilde {\cal U}(\cal H)$ of unitary operators, then the sequence 
$$1 \to U(1) \aderecha{\iota} {\cal U}({\cal H}) \aderecha{\pi} {\cal U}(\hat{\cal H})\to 1 $$ is also exact, where ${\cal U}(\hat{\cal H})$ is the image of ${\cal U}({\cal H})$ under $\pi$, and is a subgroup 
of $Aut(\hat{\cal H})$. 
${\cal U}({\cal H})$ is given the {\it strong operator topology}, that is, the smallest 
topology which makes continuous the maps $E_h: {\cal U}(\cal H)\to \cal H$, $E_h(A)
:=A(h)$, for $h \in \cal H$, and ${\cal U}(\hat{\cal H})$ is given the quotient topology relative to the projection $A \mapsto \hat{A}$. 

\

{\it Definition 2.} If $X$ is a topological space, then $f:X \to {\cal U}({\cal H})$ is said to be {\it strongly continuous} if it is continuous when ${\cal U}({\cal H})$ is given the strong operator topology. Clearly, $f$ is strongly continuous if and only if $E_h \circ f$ is continuous for each $h \in {\cal H}$. 

\

On the other hand, Bargmann's theorem (Bargmann, 1954) 
says that if $G$ is a connected and simply connected Lie group with 
2nd cohomology $H^2(Lie(G);
\R)\cong 0$, then for any {\it projective representation} of $G$ on $\cal H$ {\it i.e.} for any 
continuous group homomorphism $\rho:G \to {\cal U}(\hat{\cal H})$, there exists a 
{\it unitary representation} of $G$ on $\cal H$ {\it i.e.} a strongly continuous group homomorphism 
$\hat{\rho}:G \to {\cal U}(\cal H)$ such that $\pi \circ \hat{\rho}=\rho$ (Simms, 1968). 
$\hat{\rho}$ is called a {\it lifting } of $\rho$. Clearly, $\tilde{U}(n)$ 
satisfies the conditions of Bargmann's theorem, since by a theorem of Chevalley and Eilenberg (Chevalley and Eilenberg, 1948) the real cohomology of a compact and connected Lie group is isomorphic to the cohomology of its Lie algebra, and $H^2(U(n);\R) \cong 0$ (It\^o, 1993).
In Theorem 4 below, we give an explicit expression for the lifting $\hat{\rho}$. Moreover, in Proposition 3, we shall show that $\hat{\rho}$ is in fact a homomorphism of topological groups. 

\

The following proposition gives a useful characterization of strongly continuous maps.    

\

{\it Proposition 2.} Let $G$ be a topological group and let $f:G \to {\cal U}({\cal H})$ be a group homomorphism. Let $\bar {f} :G \times {\cal H} \to {\cal H}$ be the action given by $\bar {f}(g,h):=f(g)(h)$. Then $\bar {f}$ is continuous if and only if $f$ is strongly continuous. 

\

{\it Proof.} $\Rightarrow$) Assume that $\bar {f}$ is continuous, and consider the composition $\bar {f} \circ \alpha _h$ with $\alpha _h :G \to G \times {\cal H}$ given by $\alpha _h(g)=(g,h)$. Since $\alpha _h$ is clearly continuous, then $\bar {f} \circ \alpha _h$ is continuous. But $\bar {f} \circ \alpha _h =E_h \circ f$ {\it i.e.} $f$ is strongly continuous.

$\Leftarrow$) Assume now that $f$ is strongly continuous. Let $(g_0,h_0) \in G \times {\cal H}$. We shall prove that $\bar {f}$ is continuous at $(g_0,h_0)$. For this, let $\{(g_\lambda, h_\lambda)\}_{\lambda \in \Lambda}$ be a net in $G\times {\cal H}$ which converges to $(g_0,h_0)$. We will show that the net $\{\bar {f}(g_\lambda,h_\lambda)\}_{\lambda \in \Lambda}$ 
converges to $\bar {f}(g_0,h_0)$. 

Let $(g,h)$ be any point in $G\times {\cal H}$. Then we have the following inequality for the norm of $\bar {f} (g,h)-\bar {f} (g_0,h_0)=f(g)(h)-f(g_0)(h_0)$: 

$\vert \vert f(g)(h)-f(g_0)(h_0)\vert \vert $

=$\vert \vert f(g_0)((f(g_0))^{-1}f(g)(h)-h_0) \vert \vert $

=$\vert \vert f(g_0)(f(g_0 ^{-1}g)(h)-h_0)\vert \vert $ 

=$\vert \vert f(g_0 ^{-1}g)(h)-h_0 \vert \vert $

=$\vert \vert f(g_0 ^{-1}g)(h-h_0)+f(g_0 ^{-1}g)(h_0)-h_0 \vert \vert $

$\leq \vert \vert f(g_0 ^{-1}g)(h-h_0)\vert \vert +\vert \vert f(g_0 ^{-1}g)(h_0)-h_0 \vert \vert $

=$\vert \vert h-h_0 \vert \vert + \vert \vert f(g_0 ^{-1}g)(h_0)-h_0 \vert \vert $.   (*)

Notice that in the third and sixth steps we have used the fact that $f$ takes values in the unitary group. Now consider:

i) Let $\mu_0$ be the composition of the product in $G$ and the inclusion $G \to G \times G$ given by $g \mapsto (g_0 ^{-1},g)$. Since $\mu_0 $ is continuous and $\mu_0 (g_0)=e$, and the net $\{g_\lambda \}_{\lambda \in \Lambda} \to g_0$, then the net $\{ \mu_0 (g_\lambda)=g_0 ^{-1}g_\lambda \}_{\lambda \in \Lambda} \to e$.

ii) Since $f$ is strongly continuous, then the composition $G \aderecha{f} {\cal U}({\cal H}) \aderecha{E_{h_0}} {\cal H}$ is continuous; then $E_{h_0} \circ f \circ \alpha $ given by $g \mapsto f(g_0 ^{-1}g)(h_0)$ is also continuous. Since $\{g_\lambda \}_{\lambda \in \Lambda} \to g_0$ then $\{f(g_0 ^{-1}g_\lambda)(h_0)\}_{\lambda \in \Lambda} \to h_0$. 

iii) Let $\varepsilon >0$, then there exists $\lambda _0 \in \Lambda$ such that for all $\lambda   >\lambda _0$, $\vert \vert f(g_0 ^{-1}g_\lambda)(h_0)-h_0 \vert \vert  <\varepsilon /2$; on the other hand, since $\{h_\lambda\}_{\lambda \in \Lambda} \to h_0$, then there exists $\lambda _1 \in \Lambda$ such that for all $\lambda > \lambda _1$, $\vert \vert h_\lambda-h_0 \vert \vert < \varepsilon /2$. Since $\Lambda$ is a directed set, then there exists $\bar {\lambda} \in \Lambda$ such that $\bar {\lambda} \geq \lambda _0$ and $ \bar {\lambda}\geq \lambda _1$. Then for all $\lambda \geq \bar {\lambda}$ we have, by the inequality (*), $\vert \vert f(g_\lambda)(h_\lambda)-f(g_0)(h_0)\vert \vert \leq \vert \vert h_\lambda -h_0 \vert \vert + \vert \vert f(g_0 ^{-1}g_\lambda)(h_0)-h_0 \vert \vert < \varepsilon /2+\varepsilon /2=\varepsilon$. Therefore, $\{\bar{f}(g_\lambda,h_\lambda)\}_{\lambda \in \Lambda}$ converges to $f(g_0,h_0)$ {\it i.e.} $\bar {f}$ is continuous.   QED

\

{\it Remark 4.} We have two canonical subgroups of $SU(n)\bigcirc \hskip-11pt{\scriptstyle s}$ $\R$. The subgroup $\{(A,0)\vert A \in SU(n)\}$ which is normal, and the subgroup $\{(I,t)\vert  t \in \R \}$ which is not normal; the intersection of both subgroups is trivial so that any element $(A,t)$ can be written uniquely as the product $(A,t)=(I,t)(A,0)$. In the theory of unitary representations of semidirect products (Sternberg, 1994), one asumes that the normal subgroup is abelian. This is the case of the proper orthochronous Poincar\'e group which is the semidirect product of $\R^4$ and $SO^0(3,1)$, where $\R^4$ is normal and abelian. However, in our case, $SU(n)$ is not abelian.

\

{\it Proposition 3.} The group ${\cal U}({\cal H})$, with the strong topology, is a topological group.

\

{\it Proof.} i) Let $\mu :{\cal U}({\cal H})\times {\cal U}({\cal H}) \to {\cal U}({\cal H})$ be the product in ${\cal U}({\cal H})$ {\it i.e.} $\mu (R,S)=R \circ S$. We shall show that, for each $h \in {\cal H}$, the composition $E_h \circ \mu$ is continuous. For this, let $\{(R_\lambda,S_\lambda)\}_{\lambda \in \Lambda}$ be a net in ${\cal U}({\cal H})\times {\cal U}({\cal H})$ which converges (in the strong topology) to $(R_0,S_0)$. Let $\varepsilon >0$; since $\{R_\lambda\}_{\lambda\in \Lambda}\to R_0$ and $\{S_\lambda\}_{\lambda\in\Lambda}\to S_0$, there exist $\lambda_1$ and $\lambda_2$ in $\Lambda$ such that $\vert \vert (R_\lambda -R_0)(S_0(h))\vert \vert <\varepsilon/2$, whenever $\lambda \geq \lambda_1$, and $\vert \vert (S_\lambda -S_0)(h)\vert \vert <\varepsilon/2$, whenever $\lambda \geq \lambda_2$. Now,

$\vert \vert R_\lambda \circ S_\lambda (h)-R_0 \circ S_0 (h)\vert \vert $

=$\vert \vert (R_\lambda -R_0)(S_0(h))+R_\lambda(S_\lambda(h)-S_0(h))\vert \vert$

$\leq \vert \vert (R_\lambda -R_0)(S_0(h))\vert \vert + \vert \vert R_\lambda (S_\lambda (h)-S_0(h))\vert \vert$

=$\vert \vert (R_\lambda-R_0)(S_0(h))\vert \vert +\vert \vert (S_\lambda-S_0)(h)\vert \vert$.

Let $\bar{\lambda}$ be an element in $\Lambda$ such that $\bar{\lambda}\geq \lambda_1$ and $\bar{\lambda}\geq \lambda_2$. If $\lambda \geq \bar{\lambda}$, then $\vert \vert R_\lambda \circ S_\lambda (h)-R_0 \circ S_0 (h)\vert \vert < \varepsilon /2 +\varepsilon/2=\varepsilon$. Therefore $\{R_\lambda \circ S_\lambda(h)\}\to R_0\circ S_0(h)$ and then $E_h \circ \mu$ is continuous. 

ii) It is known that the weak and the strong topologies for the set of bounded operators on ${\cal H}$ coincide in ${\cal U}({\cal H})$. To show that the map ${\cal U}({\cal H})\to {\cal U}({\cal H})$ given by $R \mapsto R^{-1}$ is continuous, we shall use the weak topology. Let $\{R_\lambda\}_{\lambda\in \Lambda}$ be a net in ${\cal U}({\cal H})$ which converges, in the weak topology, to $R_0$. Let $h$ and $h^\prime$ be elements in ${\cal H}$. Then, given $\varepsilon >0$, there exists $\lambda_0$ in $\Lambda$ such that $\vert <h,(R_\lambda-R_0)(h^\prime)>\vert <\varepsilon$, whenever $\lambda >\lambda_0$. Now, since the operators are unitary, $R^*=R^{-1}$ and $R_0^*=R_0^{-1}$. Hence $(R_\lambda-R_0)^*=R_\lambda^*-R_0^*=R_\lambda^{-1}-R_0^{-1}$, and the inequality above can be written as $\varepsilon >\vert<h,(R_\lambda-R_0)(h^\prime)>\vert=\vert<(R_\lambda-R_0)^*(h),h^\prime>\vert=\vert <(R_\lambda^{-1}-R_0^{-1})(h),h^\prime>\vert$, whenever $\lambda >\lambda_0$. Therefore $\{R_\lambda^{-1}\}_{\lambda\in \Lambda}\to R_0^{-1}$, so that the map $R \mapsto R^{-1}$ is continuous at any point $R_0$.    QED

\

{\it Remark 5.} Naimark (Naimark, 1964) showed that the composition of bounded operators on an infinite dimensional Hilbert space is not strongly continuous. However, when the composition is restricted to the unitary operators, the above proposition shows that the composition is indeed continuous, contrary to what is claimed in Simms (Simms, 1968), p.10. 

\

{\it Corollary 1.} The action ${\cal U}({\cal H})\times {\cal H}\to {\cal H}$, given by $(A,h)\mapsto A(h)$, is continuous.

\

{\it Proof.} In Proposition 2, take $G={\cal U}({\cal H})$ and $f=id$.    QED

\section{5. Classification of projective representations}

\

In this section we give a bijection between the set of projective representations of $U(n)$ and the set of equivalence classes of certain unitary representations of $SU(n) \bigcirc \hskip-11pt{\scriptstyle s}$ $\R$. 
 
\

{\it Definition 3.} Let $\lambda_1, \lambda_2:SU(n) \bigcirc \hskip-11pt{\scriptstyle s}$ $\R \to U(1)$ be continuous homomorphisms. We define $\lambda_1 \lambda_2:SU(n)\bigcirc \hskip-11pt{\scriptstyle s}$ $\R \to U(1)$, by $(\lambda_1 \lambda_2)(a)=\lambda_1(a)\lambda_2(a)$. This map is continuous since it is the composition of the following continuous maps: $$SU(n)\bigcirc \hskip-11pt{\scriptstyle s} \ \R \aderecha{\Delta} (SU(n)\bigcirc \hskip-11pt{\scriptstyle s} \ \R)\times (SU(n)\bigcirc \hskip-11pt{\scriptstyle s} \ \R)\aderecha{\lambda_1 \times \lambda_2}U(1)\times U(1)\aderecha{\nu}U(1),$$ where $\Delta(a)=(a,a)$ and $\nu$ is the product in $U(1)$. Moreover, since $U(1)$ is abelian, the product $\nu$ is a homomorphism. Therefore $\lambda_1 \lambda_2$ is also a homomorphism, because it is a composite of homomorphisms. The rule $(\lambda_1, \lambda_2)\mapsto \lambda_1 \lambda_2$ gives a group structure to the set of continuous homomorphisms, with identity $1(a)=1$ and inverses $\lambda^{-1}(a)=\lambda (a)^{-1}$. We denote this group by $(SU(n)\bigcirc \hskip-11pt{\scriptstyle s}$ $\R)^*\equiv {\bf U_0}$. It is the {\it group of one dimensional (irreducible) representations of} $SU(n)\bigcirc \hskip-11pt{\scriptstyle s}$ $\R$. (In Proposition 9, we will show that ${\bf U_0}$ is isomorphic to $\R$.)

\

Let $\iota:\Z \to SU(n) \bigcirc \hskip-11pt{\scriptstyle s}$ $\R $ be the inclusion $\iota(k)=(I,k)$, clearly $\iota(\Z)=ker(p)$. 

\     

{\it Definition 4.} We denote by ${\cal U}(SU(n) \bigcirc \hskip-10pt{\scriptstyle s} \ \ \R)\equiv {\bf U}$ the set of unitary representations $\beta:SU(n)\bigcirc \hskip-11pt{\scriptstyle s} \ \R \to {\cal U}({\cal H})$ such that $\beta (\iota (\Z))\subset U(1)Id$, and by $hom(U(n),{\cal U}(\hat{\cal H}))\equiv {\bf P}$ the set of projective representations of $U(n)$. We define a map $Q:{\bf U}\to {\bf P}$ as follows: we associate to $\beta$, the homomorphism $Q(\beta):U(n)\to {\cal U}(\hat{{\cal H}})$, given by $Q(\beta)(p(a))=\pi \circ \beta(a)$, {\it i.e.}, $Q(\beta)$ is the homomorphism in the quotient groups, induced by $\beta$. 

\

{\it Lemma 1.} The map $Q:{\bf U} \to {\bf P}$ is well defined and it is surjective. 

\

{\it Proof.} The homomorphism $\beta$ maps $\iota(\Z)=ker(p)$ into $U(1)Id=ker(\pi)$, therefore the homomorphism $Q(\beta)$ is well defined. Since $p$ is a covering space, it is an open map so that $U(n)$ has the quotient topology. Hence $Q(\beta)$ is continuous and it is then a projective representation. 

Now let $\rho:U(n)\to {\cal U}(\hat{{\cal H}})$ be a projective representation. As we did before, we can use Bargmann's theorem for the representation $\rho \circ p$ to get a representation $\hat{\rho}:SU(n)\bigcirc \hskip-9pt{\scriptstyle s} \ \R \to {\cal U}({\cal H})$ such that $\pi \circ \hat{\rho}=\rho \circ p$, clearly $\hat{\rho}(\iota(\Z))\subset U(1)Id$. Hence $\hat{\rho}\in {\bf U}$. Since $\pi \circ \hat{\rho}=Q(\hat{\rho})\circ p$, then $\rho \circ p=Q(\hat{\rho})\circ p$ and, since $p$ is surjective, $\rho=Q(\hat{\rho})$. Therefore $Q$ is surjective.    QED

\

In order to study the map $Q$, we will define an action of the group ${\bf U_0}$ on ${\bf U}$. 

\

{\it Proposition 4.} There is a free action ${\bf U_0} \times {\bf U} \to {\bf U}$ given by $(\lambda, \beta)\mapsto \lambda \cdot \beta$, where $(\lambda \cdot \beta)(a)=\lambda(a)Id \circ \beta(a)$. 

\

{\it Proof.} i) The map $\lambda \cdot \beta $ can be written as the following composite: 
$$SU(n)\bigcirc \hskip-10pt{\scriptstyle s} \ \ \R \aderecha{\Delta} (SU(n) \bigcirc \hskip-10pt{\scriptstyle s} \ \ \R )\times (SU(n) \bigcirc \hskip-10pt{\scriptstyle s} \ \ \R )\aderecha{\lambda \times \beta}U(1)\times {\cal U}({\cal H})\aderecha{\iota \times Id}{\cal U}({\cal H})\times {\cal U}({\cal H})\aderecha{\mu} {\cal U}({\cal H}).$$ By Proposition 3, $\mu$ is continuous, therefore $\lambda \cdot \beta$ is the composition of continuous maps, so it is continuous. 

ii) Since $(\lambda \cdot \beta)(a_1 a_2)=\lambda (a_1 a_2)Id \circ \beta (a_1 a_2)=\lambda (a_1)\lambda (a_2)Id\circ \beta (a_1)\circ \beta (a_2)=\lambda (a_1)Id\circ \beta (a_1)\circ \lambda (a_2)Id \circ \beta (a_2)=(\lambda \cdot \beta)(a_1)\circ (\lambda \cdot \beta )(a_2)$, then $\lambda \cdot \beta$ is a homomorphism. 

iii) Now recall that the elements of the form $\lambda(a)Id$ belong to $ker(\pi)$ and that $\beta (ker(p))\subset ker(\pi)$. Therefore if $a\in ker(p)$, then $\pi \circ (\lambda \cdot \beta)(a)=\pi(\lambda(a)Id \circ \beta(a))=\pi (\lambda(a)Id)\circ \pi(\beta(a))=\hat{Id}$. Hence $(\lambda \cdot \beta)(ker(p))\subset ker(\pi)$. These facts show that $\lambda \cdot \beta$ is an element of ${\bf U}$. 

iv) An easy calculation, similar to ii), shows that $(\lambda, \beta) \mapsto \lambda \cdot \beta$ is an action.

v) Assume that $\lambda \cdot \beta =\beta$, then for all $a$ in $SU(n)\bigcirc \hskip-10pt{\scriptstyle s} \ \ \R$, we have that $\lambda (a)Id \circ \beta (a)=\beta (a)$, therefore $\lambda (a)Id=Id$, and $\lambda (a)=1$ {\it i.e.} $\lambda =1$. Hence the action is free.    QED

\

{\it Theorem 2.} Let $\beta_1$ and $\beta_2$ be elements in ${\bf U}$. Then $Q(\beta_1)=Q(\beta_2)$ if and only if there exists an element $\lambda$ in ${\bf U_0}$ such that $\beta_1=\lambda \cdot \beta_2$. In other words, the fibers of $Q$ are precisely the orbits  of the action of the group ${\bf U_0}$. 

\

{\it Proof.} $\Rightarrow$) Assume that $Q(\beta_1)=Q(\beta_2)$. Let $\tilde{\lambda}$ be the following composite: $$SU(n)\bigcirc \hskip-9pt{\scriptstyle s} \ \R \aderecha{\Delta}(SU(n)\bigcirc \hskip-9pt{\scriptstyle s} \ \R )\times (SU(n)\bigcirc \hskip-9pt{\scriptstyle s} \ \R )\aderecha{\beta_1 \times\beta_2}{\cal U}({\cal H})\times {\cal U}({\cal H})\aderecha{\tilde{\mu}}{\cal U}({\cal H}),$$ where $\Delta$ is the diagonal map and $\tilde{\mu}(A,B)=A\circ B^{-1}$. By Proposition 3, $\tilde{\mu}$ is continuous, hence $\tilde{\lambda}$ is continuous. Since $\pi(\beta_1(a)\circ \beta_2(a)^{-1})=\pi(\beta_1(a))\circ \pi(\beta_2(a))^{-1}=Q(\beta_1)(p(a))\circ Q(\beta_2)(p(a))^{-1}=\hat{Id}$, then $\tilde{\lambda}(a)=\beta_1(a)\circ \beta_2(a)^{-1}\in ker(\pi)=U(1)Id$. Using this fact, we have that $\tilde{\lambda}(a_1 a_2)=\beta_1(a_1 a_2)\circ \beta_2(a_1 a_2)^{-1}=\beta_1(a_1)\circ \beta_1(a_2)\circ \beta_2(a_2)^{-1}\circ \beta_2(a_1)^{-1}=\beta_1(a_1)\circ \beta_2(a_1)^{-1}\circ \beta_1(a_2)\circ \beta_2(a_2)^{-1}=\tilde{\lambda}(a_1)\circ \tilde{\lambda}(a_2)$ {\it i.e.} $\tilde{\lambda}$ is a continuous homomorphism such that $\tilde{\lambda}(SU(n)\bigcirc \hskip-11pt{\scriptstyle s}$ $\R )\subset U(1)Id$ and $\beta_1(a)=\tilde{\lambda}(a)\circ \beta_2(a)$. Since $\iota:U(1)\to {\cal U}({\cal H})$ given by $\iota (z)=zId$ is a topological embedding and a homomorphism, we have a continuous homomorphism $\lambda:SU(n)\bigcirc \hskip-11pt{\scriptstyle s} \ \ \R \to U(1)$ such that $\iota \circ \lambda=\tilde{\lambda}$, and clearly $\beta_1=\lambda \cdot \beta_2$. 

$\Leftarrow$) Assume that $\beta_1=\lambda\cdot \beta_2$. Since $\lambda(a)Id \subset ker(\pi)$, then we have that $Q(\beta_1)(p(a))=Q(\lambda \cdot \beta_2)(p(a))=\pi((\lambda \cdot \beta_2)(a))=\pi(\lambda(a)Id\circ \beta_2(a))=\pi(\beta_2(a))=Q(\beta_2)(p(a))$. Therefore $Q(\beta_1)=Q(\beta_2)$.    QED

\

{\it Corollary 2.} Let $\rho:U(n)\to {\cal U}(\hat{{\cal H}})$ be a projective representation. Then there is a bijection between the group ${\bf U_0}$ and $Q^{-1}(\{\rho \})$. 

\

{\it Proof.} By the theorem, $Q^{-1}(\{\rho \})$ coincides with the orbit of any element $\beta$  in $Q^{-1}(\{\rho \})$. But by the Proposition 4, the action is free, so that the orbit of any point is in bijective correspondence with the group ${\bf U_0}$.    QED

\

{\it Remark 6.} Once we choose an element $\beta$ in $Q^{-1}(\{\rho \})$, we have a bijection from ${\bf U_0}$ to $Q^{-1}(\{\rho \})$ given by $\lambda \mapsto \lambda \cdot \beta$. So, in general there is no canonical way to identify $Q^{-1}(\{\rho \})$ with the group ${\bf U_0}$. However, when $\rho$ is the trivial representation $\hat{c}$ given by $\hat{c}(A)=\hat{Id}$, there is a canonical element $c$ in $Q^{-1}(\{\hat{c}\})$, namely, the trivial unitary representation given by $c(a)=Id$, and in this case $Q^{-1}(\{\hat{c}\})$ can be identified with the representations ${\bf U_0}$ through $\lambda \mapsto \lambda \cdot c$. 

\

{\it Remark 7.} The bijection between the set ${\bf P}$ of projective representations of $U(n)$ and the orbits of ${\bf U}$ under the action of ${\bf U_0}$ is given by the map $\rho \mapsto Q^{-1}(\{\rho \})$. 

\

Finally, the following theorem, together with Theorem 2 above, allows a classification of the projective representations of $U(n)$ in terms of unitary representations of $SU(n)$ and one-parameter unitary groups, satisfying certain condition. 

\

{\it Theorem 3.} There is a canonical bijection between ${\bf U}$ and the set of pairs $(f_1,f_2)$, which satisfy the following conditions: 

i) $f_1:SU(n)\to {\cal U}({\cal H})$ and $f_2:\R \to {\cal U}({\cal H})$ are both continuous homomorphisms

ii) $f_2(1)\in U(1)Id$

iii) $f_1(A\cdot t)=f_2(t)^{-1}\circ f_1(A)\circ f_2(t)$, for all $A$ in $SU(n)$ and $t$ in $\R$. 

\

{\it Proof.} Let $\beta:SU(n)\bigcirc \hskip-11pt{\scriptstyle s}$ $\R \to {\cal U}({\cal H})$ be an element in ${\bf U}$. Let $\iota_1:SU(n)\to SU(n)\bigcirc \hskip-10pt{\scriptstyle s}$ $\R$ and $\iota_2:\R \to SU(n)\bigcirc \hskip-10pt{\scriptstyle s}$ $\R$ be the canonical inclusions. Then we associate to $\beta$ the pair $(\beta \circ \iota_1,\beta \circ \iota_2)$; clearly both are continuous homomorphisms. Since $\beta(\iota(\Z))\subset U(1)Id$, then $\beta \circ \iota_2 (1)=\beta \circ \iota (1)$ is an element of $U(1)Id$. Since $\beta$ is a homomorphism, and any element $(A,t)$ in $SU(n)\bigcirc \hskip-11pt{\scriptstyle s}$ $\R$ can be written as $(A,t)=(I,t)(A,0)=\iota_2(t)\iota_1(A)$, then $\beta((A,t)(A^\prime,t^\prime))=\beta \circ \iota_2(t)\circ \beta \circ \iota_1(A)\circ \beta \circ \iota_2(t^\prime)\circ \beta \circ \iota_1(A^\prime)$; but $\beta ((A,t)(A^\prime,t^\prime))=\beta((A\cdot t^\prime)A^\prime, t+t^\prime)=\beta(\iota_2(t+t^\prime)\iota_1((A\cdot t^\prime)A^\prime))=\beta \circ \iota_2(t)\circ \beta \circ \iota_2(t^\prime)\circ \beta \circ \iota_1(A\cdot t^\prime)\circ \beta \circ \iota_1(A^\prime)$, then $\beta \circ \iota_1(A\cdot t^\prime)=\beta \circ \iota_2(t^\prime)^{-1}\circ \beta \circ \iota_1(A)\circ \beta \circ \iota_2(t^\prime)$. 

Conversely, given a pair $(f_1,f_2)$ satisfying i), ii) and iii), we define $f:SU(n)\bigcirc \hskip-11pt{\scriptstyle s}$ $\R \to {\cal U}({\cal H})$ by $f(A,t)=f_2(t)\circ f_1(A)$. This map can be written as the composite $$SU(n)\bigcirc \hskip-9pt{\scriptstyle s} \ \R \aderecha{f_1 \times f_2}{\cal U}({\cal H})\times {\cal U}({\cal H})\aderecha{s}{\cal U}({\cal H})\times {\cal U}({\cal H})\aderecha{\mu}{\cal U}({\cal H})$$ where $s(A,B)=(B,A)$. By Proposition 3, $\mu$ is continuous, hence $f$ is continuous. Now, using property iii) we can write: $f((A,t)(A^\prime,t^\prime))=f((A\cdot t^\prime)A^\prime,t+t^\prime)=f_2(t)\circ f_2(t^\prime)\circ f_1(A\cdot t^\prime)\circ f_1(A^\prime)=f_2(t)\circ f_2(t^\prime)\circ f_2(t^\prime)^{-1}\circ f_1(A)\circ f_2(t^\prime)\circ f_1(A^\prime)=f_2(t)\circ f_1(A)\circ f_2(t^\prime)\circ f_1(A^\prime)=f(A,t)\circ f(A^\prime,t^\prime)$ {\it i.e.} $f$ is a homomorphism; and $f(I,n)=f_2(n)\circ f_1(I)=f_2(n)\circ Id=f_2(n)$, which is an element of $U(1)Id$, by property ii). 

Finally, one construction is the inverse of the other. Indeed, let $f_1=\beta \circ \iota_1$ and $f_2=\beta \circ \iota_2$, then $f(A,t)=\beta \circ \iota_2(t)\circ \beta \circ \iota_1(A)=\beta (\iota_2(t))\circ \beta(\iota_1(A))=\beta(I,t)\circ \beta(A,0)=\beta((I,t)(A,0))=\beta(A,t)$ {\it i.e.} $f=\beta$; conversely, given $f$ constructed  from the pair $(f_1,f_2)$, define the pair $(f\circ \iota_1, f\circ \iota_2)$, then $f\circ \iota_1(A)=f(A,0)=f_2(0)\circ f_1(A)=Id\circ f_1(A)=f_1(A)$ and $f\circ \iota_2(t)=f(I,t)=f_2(t)\circ f_1(I)=f_2(t)\circ Id=f_2(t)$ {\it i.e.} $f\circ \iota_1=f_1$ and $f\circ \iota_2=f_2$.    QED

\section{6. Unitary representations associated to a projective representation}

\

In this section we shall construct all the unitary representations of $\tilde{U}(n)$ associated to a projective representation of $U(n)$.

\

{\it Proposition 5.} Let $G$ be a connected simple Lie group and let $K$ be a Lie group such that $dim$ $G>dim$ $K$. Let $\gamma:G\to K$ be a continuous homomorphism. Then $\gamma$ is trivial {\it i.e.}, $\gamma(g)=e$, for all $g\in G$.

\

{\it Proof.} By (Warner, 1983), any continuous homomorphism between Lie groups is smooth, so we can assume that $\gamma$ is smooth. Consider the differential of $\gamma$ at the identity, $d\gamma:\cal G \to \cal K$, where $\cal G$ and $\cal K$ are respectively the Lie algebras of $G$ and $K$. Since $G$ is simple, $ker(d\gamma)$ is either 0 or $\cal G$. Suppose that $ker(d\gamma)=0$, then $dim$ ${\cal G}=dim$ $d\gamma({\cal G})\leq dim$ ${\cal K}$ which is a contradiction because $dim$ ${\cal G} > dim$ ${\cal K}$. Therefore $ker(d\gamma)={\cal G}$ {\it i.e.}, $d\gamma \equiv 0$, and since $G$ is connected, by (Warner, 1983), $\gamma$ is trivial.    QED

\

{\it Remark 8.} Let $\R^*=hom(\R,U(1))$ be the group of continuous homomorphisms $\chi:\R \to U(1)$, where $(\chi_1 \chi_2)(t)=\chi_1(t)\chi_2(t)$. Then there is an isomorphism from $\R$ to $\R^*$ given by $r\mapsto \chi_r$, where $\chi_r(t)=e^{2\pi irt}$. This can be shown as follows. Recall that if $G$ is a Lie group, then we have a function from its Lie algebra $\cal G$ to the set of one-parameter subgroups of $G$, given by $X\mapsto exp(tX)$. Let $\psi:\R \to G$ be a one-parameter subgroup. Since $\psi$ is the maximal integral curve of the left invariant vector    field $X=\dot{\psi}(0)$ starting from $e$, then $\psi(t)=exp(tX)$. Therefore the function $X\mapsto exp(tX)$ is a bijection. Taking $G=U(1)$ and using the fact that continuous homomorphisms between Lie groups are smooth (Warner, 1983), we get a bijection between $i\R$ and $\R^*$, which in this case is clearly an isomorphism of groups. If we also consider the isomorphism from $\R$ to $i\R$ given by $r\mapsto2\pi ir$, then we get an isomorphism $\R \cong \R^*$ mapping $r$ to $\chi_r$.  

\

{\it Definition 5.} Let $G$ be a topological group. We denote by $hom(G,{\cal U}({\cal H}))$ the set of continuous homomorphisms from $G$ to ${\cal U}({\cal H})$, {\it i.e.}, the set of unitary representations of $G$, and by $hom(G,{\cal U}(\hat{\cal H}))$ the set of continuous homomorphisms from $G$ to ${\cal U}(\hat{\cal H})$ {\it i.e.}, the set of projective representations of $G$. 

\

{\it Proposition 6.} Let $G$ be a connected, simply connected, simple Lie group such that $H^2({\cal G},\R)=0$. Then the map $\pi:{\cal U}({\cal H})\to {\cal U}(\hat{\cal H})$ induces a bijection $\pi_*:hom(G,{\cal U}({\cal H}))\to hom(G,{\cal U}(\hat{\cal H}))$, where $\pi_*(\beta)=\pi \circ \beta$. 

\

{\it Proof.} By Bargmann's theorem, given $\rho\in hom(G,{\cal U}(\hat{\cal H}))$, there is an element $\tilde{\rho}$ in $hom(G,{\cal U}({\cal H}))$ such that $\pi \circ \tilde{\rho}=\rho$, {\it i.e.}, $\pi_*(\tilde{\rho})=\rho$. Therefore $\pi_*$ is surjective. To show that $\pi_*$ is injective, consider $\beta_1$ and $\beta_2$ in $hom(G,{\cal U}({\cal H}))$ and assume that $\pi_*(\beta_1)=\pi_*(\beta_2)$. Then for all elements $g$ in $G$ we have that $\pi(\beta_1(g))=\pi(\beta_2(g))$. Hence $\pi(\beta_1(g)\beta_2(g)^{-1})=\hat{Id}$. Define $\gamma:G\to {\cal U}({\cal H})$ by $\gamma(g)=\beta_1(g)\beta_2(g)^{-1}$. By Proposition 3, ${\cal U}({\cal H})$ is a topological group, so that $\gamma$ is continuous. Furthermore, $\gamma(G)\subset U(1)Id \cong U(1)$, whose elements commute with any element in ${\cal U}({\cal H})$. Then $\gamma(g_1 g_2)=\beta_1(g_1 g_2)\beta_2(g_1 g_2)^{-1}=\beta_1(g_1)\beta_1(g_2)\beta_2(g_2)^{-1}\beta_2(g_1)^{-1}=\beta_1(g_1)\beta_2(g_1)^{-1}\beta_1(g_2)\beta_2(g_2)^{-1}=\gamma(g_1)\gamma(g_2)$. Therefore $\gamma:G\to U(1)$ is a continuous homomorphism, so by Proposition 5, $\gamma$ is trivial. Then $1=\gamma(g)=\beta_1(g)\beta_2(g)^{-1}$, for all $g$ in $G$; thus $\beta_1=\beta_2$ and $\pi_*$ is injective.    QED

\

{\it Definition 6.} We define an action $\cdot:\R\times hom(\R,{\cal U}({\cal H}))\to hom(\R,{\cal U}({\cal H}))$ by $(r\cdot \lambda)(t)=e^{2\pi irt}Id \circ \lambda(t)$. 

\

{\it Lemma 2.} The action is well defined and free.

\

{\it Proof.} i) $(r\cdot \lambda)(t)$ is in ${\cal U}({\cal H})$ since it is the composition of elements of ${\cal U}({\cal H})$. 

ii) $r\cdot \lambda:\R \to {\cal U}({\cal H})$ is continuous since it is the composite: $$\R \to U(1)\times {\cal U}({\cal H})\to {\cal U}({\cal H})\times {\cal U}({\cal H})\to {\cal U}({\cal H}),$$ $$t\mapsto (e^{2\pi irt},\lambda(t))\mapsto (e^{2\pi irt}Id,\lambda(t))\mapsto e^{2\pi irt}Id\circ \lambda(t).$$

iii) $r\cdot \lambda$ is a homomorphism because $(r\cdot \lambda)(t_1+t_2)=e^{2\pi ir(t_1+t_2)}Id\circ \lambda (t_1+t_2)=e^{2\pi irt_1}Id\circ \lambda(t_1)\circ e^{2\pi irt_2}Id\circ \lambda(t_2)=(r\cdot \lambda)(t_1)\circ (r\cdot \lambda)(t_2)$.

iv) It is an action because $(0\cdot \lambda)(t)=\lambda(t)$ and $(r_1+r_2)\cdot \lambda(t)=e^{2\pi i(r_1+r_2)t}Id\circ \lambda(t)=(e^{2\pi ir_1t}e^{2\pi ir_2t})Id\circ \lambda(t)=e^{2\pi ir_1t}Id\circ e^{2\pi ir_2t}Id\circ \lambda(t)=r_1\cdot (r_2\cdot \lambda)(t)$. 

v) Assume that $r\cdot \lambda=\lambda$, then $(r\cdot \lambda)(t)=e^{2\pi irt}Id\circ \lambda(t)=\lambda(t)$ for all $t$ in $\R$. This implies that $e^{2\pi irt}=1$ for all $t$ in $\R$, therefore $r=0$.    QED

\

{\it Proposition 7.} Let $\lambda_1, \lambda_2:\R \to {\cal U}({\cal H})$ be continuous homomorphisms. Then $\pi \circ \lambda_1=\pi \circ \lambda_2$ if and only if there exists $r$ in $\R$ such that $\lambda_1=r\cdot \lambda_2$. 

\

{\it Proof.} Assume that $r\cdot \lambda_1=\lambda_2$. Then $\lambda_2(t)=e^{2\pi irt}Id\circ \lambda_1(t)$ and $\pi(\lambda_2(t))=\pi(e^{2\pi irt}Id)\circ \pi(\lambda_1(t))=\pi(\lambda_1(t))$. Therefore, $\pi \circ \lambda_1=\pi \circ \lambda_2$. 

Assume that $\pi \circ \lambda_1=\pi \circ \lambda_2$. Then $\pi(\lambda_1(t)\lambda_2(t)^{-1})=\hat{Id}$ and we have that $\lambda_1(t)\lambda_2(t)^{-1}$ is in $ker(\pi)=U(1)Id$, for all $t$ in $\R$. Define $\psi:\R\to U(1)$ by $\psi(t)=\lambda_1(t)\lambda_2(t)^{-1}$. By Proposition 3, ${\cal U}({\cal H})$ is a topological group, so $\psi $ is continuous. Furthermore, $\psi(t_1+t_2)=\lambda_1(t_1+t_2)\lambda_2(t_1+t_2)^{-1}=\lambda_1(t_1)\circ \lambda_1(t_2)\circ \lambda_2(t_2)^{-1}\circ \lambda_2(t_1)^{-1}=\lambda_1(t_1)\circ \lambda_2(t_1)^{-1}\circ \lambda_1(t_2)\circ \lambda_2(t_2)^{-1}=\psi(t_1)\circ \psi(t_2)$. Therefore, $\psi$ is a continuous homomorphism. By Remark 8, there exists a unique real number $r$ such that $\psi(t)=e^{2\pi irt}$. Hence $\lambda_1(t)=\psi(t)\lambda_2(t)=e^{2\pi irt}Id\circ \lambda_2(t)=(r\cdot \lambda_2)(t)$, so that $\lambda_1=r\cdot \lambda_2$.    QED

\

{\it Proposition 8.} The map $\pi:{\cal U}({\cal H})\to {\cal U}(\hat{\cal H})$ induces a surjection $$\pi_*:hom(\R,{\cal U}({\cal H}))\to hom(\R,{\cal U}(\hat{\cal H}))$$ given by $\pi_*(\lambda)=\pi \circ \lambda$, whose fibers are in one to one correspondence with $\R$.

\ 

{\it Proof.} Since $\R$ satisfies the hypotesis of Bargmann's theorem, given any one-parameter subgroup $\delta:\R\to {\cal U}(\hat{\cal H})$, there exists $\tilde{\delta}$ in $hom(\R,{\cal U}({\cal H}))$ such that $\pi \circ \tilde{\delta}=\delta$, {\it i.e.}, $\pi_*(\tilde{\delta})=\delta$, so $\pi_*$ is surjective.

By Proposition 7, given $\delta$ in $hom(R,{\cal U}(\hat{\cal H}))$, $\pi_*^{-1}(\{\delta\})$ coincides with an orbit of the action of $\R$ on $hom(\R, {\cal U}({\cal H}))$. By Lemma 2, this action is free, so each orbit is in one to one correspondence with the group $\R$.    QED

\

{\it Remark 9.} Let $\lambda$ be in $hom(\R,{\cal U}({\cal H}))$. Since $\lambda $ is a strongly continuous one-parameter unitary group, then, by Stone's theorem (Reed and Simon, 1972), there exists a unique Hermitian (though not necessarily bounded) operator $H$ on $\cal H$ such that $\lambda (t)=e^{iHt}$.

\

{\it Proposition 9.} ${\bf U_0}$=$\{f:SU(n)\bigcirc \hskip-11pt{\scriptstyle s}$ $\R \to U(1)|f$  is a continuous homomorphism$\}$ is isomorphic to $\R$. 

\

{\it Proof.} We will give first an isomorphism between ${\bf U_0}$ and $\R^*=\{\chi:\R \to U(1)|\chi$ is a continuous homomorphism$\}$. Let $f$ be in ${\bf U_0}$ and consider $f\circ \iota_1:SU(n)\to U(1)$. Then by Proposition 5 we know that $f\circ \iota_1$ is constant. So we define a function $F:{\bf U_0}\to \R^*$ by $F(f)=f\circ \iota_2:\R \to U(1)$. Given $\chi$ in $\R^*$, consider the map $\bar{\chi}:SU(n)\bigcirc \hskip-9pt{\scriptstyle s}$ $\R \to U(1)$ given by $\bar{\chi}(A,t)=\chi(t)$. Since $\bar{\chi}=\chi \circ \pi_2$, where $\pi_2:SU(n)\bigcirc \hskip-11pt{\scriptstyle s}$ $\R \to \R$ is the projection, and both $\chi$ and $\pi_2$ are continuous homomorphisms, then $\bar{\chi}$ is in ${\bf U_0}$. Furthermore, $F(\bar{\chi})=\bar{\chi}\circ \iota_2=\chi$, so $F$ is surjective. 

Now let $f_1, f_2$ be in ${\bf U_0}$; then $F(f_1 f_2)(t)=(f_1 f_2)\circ \iota_2(t)=f_1(\iota_2(t))f_2(\iota_2(t))=f_1\circ \iota_2(t)f_2\circ \iota_2(t)=F(f_1)(t)F(f_2)(t)$. Hence $F$ is a homomorphism. 

Let $f$ be in ${\bf U_0}$ and assume that $F(f)$ is trivial, {\it i.e.}, $F(f)(t)=1$ for all $t$ in $\R$. Any element $(A,t)$ in $SU(n)\bigcirc \hskip-9pt{\scriptstyle s}$ $\R$ can be written as $(A,t)=(I,t)(A,0)=\iota_2(t)\iota_1(A)$, therefore $f(A,t)=f\circ \iota_2(t)f\circ \iota_1(A)$. Since $F(f)=f\circ \iota_2$ is trivial, then $f(A,t)=f\circ \iota_1(A)$, but we saw  that $f\circ \iota_1$ is also trivial, so $f$ is trivial, and then $F$ is injective. Therefore $F$ is an isomorphism. 

Finally, by Remark 8, there is an isomorphism from $\R$ to $\R^*$ given by $r\mapsto \chi_r$, where $\chi_r(t)=e^{2\pi irt}$. Using both isomorphisms we get an isomorphism from $\R$ to ${\bf U_0}$ given by $r\mapsto f_r:SU(n)\bigcirc \hskip-11pt{\scriptstyle s}$ $\R \to U(1)$, where $f_r(A,t)=e^{2\pi irt}$.    QED

\

{\it Theorem 4.} Let $\rho:U(n)\to {\cal U}(\hat{\cal H})$ be a projective representation. Then there is a bijection between $Q^{-1}(\{\rho \})$ and the set $\{(\ro \circ \iota_1,r\cdot(\ro \circ \iota_2)|r\in \R \}$, where $\ro:SU(n)\bigcirc \hskip-8pt{\scriptstyle s}$ $\R \to {\cal U}({\cal H})$ is any fixed representation in $Q^{-1}(\{\rho\})$ and where $r\cdot (\ro \circ \iota_2)(t)=e^{2\pi irt}Id\circ \ro (Id,t)$. The bijection is given by $(\ro \circ \iota_1,r\cdot(\ro \circ \iota_2))\mapsto \hat{\rho}:SU(n)\bigcirc \hskip-11pt{\scriptstyle s}$ $\R \to {\cal U}({\cal H})$, where $\hat{\rho}(A,t)=e^{2\pi irt}Id\circ \ro (Id,t)\circ \ro(A,0)$.

\

{\it Proof.} We first recall that the existence of $\ro$ is given by Bargmann's theorem. Now, by  Theorem 3, there is a canonical bijection between $Q^{-1}(\{\rho\})$ and the set $\cal A$ of pairs $(f_1,f_2)$ which satisfy the conditions i), ii), iii) plus the following condition iv) to ensure that the elements belong to $Q^{-1}(\{\rho\})$:

iv) $Q(f)=\rho$, where $f(A,t)=f_2(t)\circ f_1(A)$

Let $\cal B$ be the set of pairs $\{(\ro \circ \iota_1,r\cdot (\ro \circ \iota_2)|r\in \R\}$ in the statement of the theorem. We will show that $\cal A =\cal B$. 

Let $(\ro \circ \iota_1,r\cdot(\ro \circ \iota_2))$ be in $\cal B$. Clearly $\ro \circ \iota_1$ is a continuous homomorphism, and the same is true for $r\cdot (\ro \circ \iota_2)$ because ${\cal U}({\cal H})$ is a topological group, so i) is satisfied. 

Since $\ro$ is an element of ${\bf U}$, then $\ro \circ \iota_2(1)=\ro (Id,1)$ belongs to $U(1)Id$, hence $r\cdot (\ro \circ \iota_2)(1)=e^{2\pi ir}Id\circ \ro (Id,1)$ is in $U(1)Id$, so ii) is satisfied. 

By Theorem 3, $\ro \circ \iota_1(A\cdot t)=\ro \circ \iota_2(t)^{-1}\circ \ro \circ \iota_1 (A)\circ \ro \circ \iota_2(t)$. Since the elements of the form $e^{2\pi irt}Id$ commute with any operator, the right hand side is equal to $\ro \circ \iota_2(t)^{-1}\circ e^{-2\pi irt}Id \circ \ro \circ \iota_1(A)\circ e^{2\pi irt}Id \circ \ro \circ \iota_2(t)=r\cdot (\ro \circ \iota_2)(t)^{-1}\circ \ro \circ \iota_1(A)\circ r\cdot (\ro \circ \iota_2)(t)$. Therefore the pair in $\cal B$ satisfies iii). 

Condition iv) is also fulfilled since $Q(\hat{\rho})(p(A,t))=\pi \circ \hat{\rho}(A,t)=\pi(e^{2\pi irt}Id)\circ \pi(\ro (Id,t))\circ \pi(\ro (A,0))=\rho (p(Id,t))\circ \rho (p(A,0))=\rho (p((Id,t)(A,0)))=\rho (p(A,t))$. Therefore $\cal B \subset \cal A$.

Conversely, let $(f_1,f_2)$ be an element in $\cal A$. Consider $\rho \circ \iota:SU(n)\to {\cal U}(\hat{\cal H})$, since $\pi \circ \ro \circ \iota_1=\rho \circ p\circ \iota_1=\rho \circ \iota$, then $\ro \circ \iota_1$ is a lifting of $\rho \circ \iota$. Let $A$ be in $SU(n)$, then  by iv) we have that $Q(f)(\iota (A))=Q(f)(p(A,0))=\pi (f(A,0))=\pi (f_2(0)\circ f_1(A))=\pi (f_1(A))=\rho \circ \iota (A)$, {\it i.e.}, $f_1$ is also a lifting of $\rho \circ \iota$. By Proposition 6, $f_1=\ro \circ \iota$. Now consider $\rho \circ \alpha :\R \to {\cal U}(\hat{\cal H})$, since $\pi \circ \ro \circ \iota_2=\rho \circ p \circ \iota_2=\rho \circ \alpha$, then $\ro \circ \iota_2$ is a lifting of $\rho \circ \alpha$. By iv) we have that $Q(f)(\alpha (t))=Q(f)(p(Id,t))=\pi (f(Id,t))=\pi (f_2(t) \circ f_1(Id))=\pi (f_2(t))=\rho (\alpha (t))$, {\it i.e.}, $f_2$ is also a lifting of $\rho \circ \alpha$. By Proposition 7, there exists $r$ in $\R$ such that $f_2=r\cdot (\ro \circ \iota_2)$. Therefore $(f_1,f_2)=(\ro \circ \iota, r\cdot (\ro \circ \iota_2))$ and then $\cal A \subset \cal B$. 

Finally, by Theorem 3, the bijection between $\cal A = \cal B$ and $Q^{-1}(\{\rho\})$ is given by mapping a pair $(f_1,f_2)$ to $f$, where $f(A,t)=f_2(t)\circ f_1(A)$, therefore $(\ro \circ \iota_1,r\cdot (\ro \circ \iota_2))$ is mapped to $\hat{\rho}$ where $\hat{\rho}(A,t)=e^{2\pi irt}Id \circ \ro (Id,t)\circ \ro (A,0)$.    QED

\section{REFERENCES}

\

\noindent Aguilar, M. A., Gitler, S., and Prieto, C. (1998). {\it Topolog\'\i a Algebraica: Un Enfoque Homot\'opico}, McGraw Hill, M\'exico. 

\

\noindent Bargmann, V. (1954). On unitary ray representations of continuous groups, {\it 
Ann. Math.} {\bf 59}, 1-46.

\

\noindent Bargmann, V. and Moshinsky, M. (1960). Group theory of harmonic oscillators. Part I: The Collective Modes, {\it Nuclear Physics} {\bf 18}, 697-712.

\

\noindent Belinfante, J. G. F. and Kolman, B. (1972). {\it A Survey of Lie Groups and Lie Algebras with Applications and Computational Methods}, Society for Industrial and Applied Mathematics, Philadelphia. 

\

\noindent Chevalley, C. and Eilenberg, S. (1948). Cohomology theory of Lie groups and Lie algebras, {\it Trans. Am. Math. Soc.} {\bf 63}, 85-124.

\

\noindent Cornwell, J. F. (1984). {\it Group Theory in Physics. Volume II}, Academic Press, London. 

\

\noindent Feynman, R. P. and Hibbs, A. R. (1965). {\it Quantum Mechanics and Path 
Integrals}, McGraw-Hill, New York. 

\

\noindent Fulton, W. and Harris, J. (1991). {\it Representation Theory, A First Course}, 
Springer-Verlag, New York.

\

\noindent It\^o, K. ed. (1993). {\it Encyclopedic Dictionary of Mathematics}, The MIT
Press, Cambridge, Mass. 

\

\noindent Naimark, M. A. (1964). {\it Normed Rings}, Noordhof.

\

\noindent Reed, M. and Simon, B. (1972). {\it Methods of Modern Mathematical Physics, I: Functional Analysis}, Academic Press, New York.

\

\noindent Simms, D. J. (1968). {\it Lie Groups and Quantum Mechanics}, Lecture Notes in 
Mathematics {\bf 52}, Springer-Verlag, Berlin.

\

\noindent Sternberg, S. (1994). {\it Group Theory and Physics}, Cambridge University Press, Cambridge. 

\

\noindent Warner, F. W. (1983). {\it Foundations of Differentiable Manifolds and Lie 
Groups}, Springer-Verlag, New York. 

\

\noindent Wigner, E. (1931). {\it Group Theory and its Application to the Quantum Mechanics 
of Atomic Spectra}, Vieweg and Son, Brunswick, Germany (in German); English transl., 
Academic Press, New York, 1959.

\

\end

The {\it suspension} of a topological space $X$ is the quotient space given  
by 
$$SX={X \times I \over {X\times \{0\},X\times \{1\}}}=\{[x,t]\}_{(x,t)\in X 
\times I}$$
with 
$$[x,t]=\bigg\{\matrix{\{(x,t)\},t \in (0,1) \cr
X\times \{0\}, t=0 \cr X\times \{1\}, t=1. \cr}$$
This means that in the product $X\times I$, $X\times 
\{0\}$ has been identified to one point, and $X\times \{1\}$ has been 
identified to another point. Intuitively, it is clear that $SS^0\cong S^1$, 
$SS^1\cong S^2$, ..., $SS^{n-1}\cong S^n$. The suspension of a continuous  
function is defined by $Sf([x,t])=[f(x),t]$, which satisfies the {\it 
functorial} properties $Sid_X=id_{SX}$ and $S$ $g\circ f=Sg\circ Sf$ if 
$f:X\to Y$ and $g:Y \to Z$. If $p_Z:Z \times I \to SZ$ is the projection 
$p(z,t)=[z,t]$ then the following diagram commutes:
$$\matrix{X & \aderecha{f}\ & Y \cr \iota_0 \downarrow & & \downarrow 
\iota_0 \cr X \times I & \aderecha{f \times id}\ & Y \times I \cr 
p_X \downarrow & & \downarrow p_Y \cr SX & \aderecha{Sf}\ & SY \cr}.$$
If $H:X \times I \to Y$ is 
a homotopy between $h_0$ and $h_1$ then $SH:SX \times I 
\to  SY$ given by $SH([x,t],t^\prime)=[H(x,t^\prime),t]$ {\it i.e.} 
$(SH)_{t^\prime}=SH_{t^\prime}$ is a homotopy between $Sh_0$ and $Sh_1$.  
$SH$ is called the {\it suspension of the homotopy}. Then there is a well 
defined function between homotopy classes of maps $S:[X,Y] \to [SX,SY]$, 
$[f] \to S([f]):=[Sf]$.

If $X$ is a pointed space with base point $x_0$, then the 
{\it reduced suspension} of $X$, $S_rX$ is defined by
$$S_rX={X \times I \over X \times \{0\} \cup X \times \{1\} \cup \{x_0\}
\times I}$$ {\it i.e.} all the points in $X \times \{0\}$, $X \times 
\{1\}$ and $\{x_0\}$ are identified to one point. In this case its 
elements are given by 
$$[x,t]= \bigg\{\matrix{X\times \{0\}=X\times \{1\}=\{x_0\} 
\times I, \matrix{x=x_0,\ all \ t \in I \cr t=0 \ or \ 1,\ all \ x \in X 
\cr} \cr \{(x,t)\}, t \in (0,1) \ and \ x \neq x_0. \cr}$$
$\tilde{x}_0=X \times \{0\}$ is the base point of $S_rX$. If $f:X \to Y$  
respects the base points {\it i.e.} if $f(x_0)=y_0$, then $S_rf(\tilde{x}
_0)=y_0$, and if $h_0 \matrix{H \cr \sim \cr}h_1 \ (rel \ x_0)$ then 
$Sh_0 \matrix{SH \cr \sim \cr}Sh_1 \ (rel \ \tilde{x}_0)$. ($rel \ x_0$ 
means that the homotopy $H$ respects the base point.) Also, there 
is a homeomorphism $\varphi^{-1}_{n+1}:S_rS^n \to S^{n+1}$ given by
$$\varphi ^{-1}_{n+1}([\vec {x},t])=\bigg\{\matrix{p^{-1}_-(2t\vec {x}+(1-2t)
\vec{x}_0), \ t \in [0,1/2] \cr p^{-1}_+((2-2t)\vec{x}+(2t-1)\vec{x}_0), 
\ t \in [1/2,1] \cr}$$ 
where : $\vec {x}_0=(1,0,...,0) \in S^{n+1}=\{(x_1,...,x_{n+2}) \vert 
\sum\limits^{n+2}_{i=1} x^2_i=1 \} \subset \R^{n+2}$ is the base point, 
$S^n=\{\vec {x}\in S^{n+1} \vert x_{n+2}=0\}$, and if $H_+=\{\vec{x} 
\in S^{n+1} \vert x_{n+2} \geq 0\}$, $H_-=\{\vec {x} \in S^{n+1} 
\vert x_{n+2} \leq 0\}$ and $D^{n+1}=\{(x_1,...,x_{n+1},0) \vert 
\sum\limits ^{n+1}_{i=1} x_i^2 \leq 1 \}$ then $p_{+(-)}
:H_{+(-)} \to D^{n+1}$ are the homeomorphisms given by $(
x_1,...,x_{n+2})\mapsto (x_1,...,x_{n+1},0)$ with inverses $(x_1,..., 
x_{n+1},0) \mapsto (x_1,...,x_{n+1},+(-) \sqrt{1-\sum
\limits ^{n+1}_{i=1} x_i^2 })$ respectively (Spanier, 1966). In particular, 
$$\tilde{\vec{x}}_0=S^n \times \{0\}=S^n \times \{1\}=\{\vec{x}_0\} \times I
\aderecha{\varphi^{-1}_{n+1}}\ \vec{x}_0$$ and if $\vec{x}\neq \vec{x}_0$ 
then $[\vec{x},1/2]=\{(\vec{x},1/2)\}\aderecha{\varphi^{-1}_{n+1}}\ \vec{x}.$ 

The inverse homeomorphism is given by the following formulae: $\vec{x}_0 
\mapsto \tilde {\vec{x}}_0$, if $\vec {x} \in S^n$ and $\vec{x} \neq 
\vec{x}_0$ then $\vec{x} \mapsto [\vec{x},1/2]=\{(\vec{x},1/2)\}$, 
$(0,...,0,1)=N \ (north \ pole) \mapsto [-\vec{x}_0,3/4]=
\{(-\vec{x},3/4)\}$, 
$(0,...,-1)=S \ (south \ pole) \mapsto [-\vec{x}_0,1/4]=
\{(-\vec{x}_0,1/4)\}$, and if        
$\vec{x} \in H_{+(-)} \ \setminus \ S^n$, $\vec{x} \neq S,N$, then 
$$\varphi_{n+1}(\vec{x})=[\vec{z}(\vec{x}),t_{+(-)}(\vec{x})]=
\{(\vec{z}(\vec{x}),t_{+(-)}(\vec{x}))\}$$ with 
$$\vec{z}(\vec{x})={(2(1-x_1)x_1-x^2_{n+2},
2(1-x_1)x_2,...,2(1-x_1)x_{n+1}) \over {2(1-x_1)-x^2_{n+2}}}$$ 
and $t_{+(-)}(\vec{x})=1/2+(-){x^2_{n+2}\over {4(1-x_1)}}$.

\subsection{3.2. $SU(3)$ from the Hopf map $h$}

Let $G$ be a path connected topological group. Then the set of 
isomorphism classes of principal $G$-bundles over the $n$-sphere 
$k_G(S^n)$ is in one-to-one correspondence with $\Pi _{n-1}(G)$ 
(Steenrod, 1951). This can be understood from the fact that the 
$n$-sphere can be covered by two open sets $U_1, U_2$, which are 
homeomorphic to $n$-balls and contain $S^{n-1}$, and the fact that 
any bundle over an $n$-ball is trivial. Using these trivializations 
there is only one transition function $g_{12}:U_1 \cap U_2 \to G$, for a 
bundle $\xi$. Then we associate to $\xi$ the map $g_{12}\mid _
{S^{n-1}}:S^{n-1}\to G$, called the characteristic map of $\xi$. 
Therefore, if the characteristic maps of two bundles are in the same 
homotopy class, then the corresponding bundles are 
isomorphic, and a bundle is trivial if and only if its characteristic 
map is null-homotopic. Notice that in our construction of 
the local charts for $SU(3)$ we have used a different trivialization. In 
the following we shall consider the bundles $SU(n-1) \to SU(n) \aderecha
{\pi_n}\ S^{2n-1}$ and call $g_n:S^{2n-2}\to SU(n-1)$ 
the corresponding characteristic maps.

To study the case $n$=3 we need the following

{\it Proposition.} The sucessive suspensions of the Hopf map, $S_rh:
S^4 \to S^3$, $S^2_rh:S^5\to S^4$,... are essential (Steenrod and 
Epstein, 1962).

As a consequence, we have the 

{\it Proposition.} $SU(3)$ is determined by the suspension of the Hopf map. 

{\it Proof.} For $n=3$, $k_{SU(2)}(S^5) \cong [S^4,S^3] \cong \Pi_4(S^3) \cong 
\Z_2=\{0,1\}$ and $g_3:S^4 \to S^3$. 
By the proposition above $S_r h$ is essential. To see that 
$g_3$ is also essential we will show that the bundle $SU(3) \aderecha{
\pi _3} S^5$ is not trivial. By  
(EDM, 1993), $\Pi _4(SU(3)) \cong 0$, on the other hand 
$\Pi _4(S^5 \times SU(2)) \cong 
\Pi _4(S^5)\times \Pi _4(S^3) \cong 0 \times \Z_2 \cong \Z _2$. Hence $SU(3)$ is not 
isomorphic to the trivial bundle. Since $\Pi _4(S^3) \cong \Z _2$ we have 
that $[g_3]=[S_r h]$.   QED

\section{4. The case of $SU(n)$}

\subsection{4.1. $SU(4)$}

For $n=4$, $k_{SU(3)}(S^7)\cong [S^6,SU(3)]\cong \Pi_6(SU(3))\cong 
\Z_6$ which has two generators. 
This means that up to isomorphism there are five nontrivial $SU(3)$-
bundles over $S^7$, one of them being $SU(4)$ since $\Pi _6(SU(4)) \cong 
0$ and $\Pi _6(S^7 \times SU(3)) \cong \Pi _6(S^7) \times \Pi _6(SU(3)) 
\cong \Z_6$. Let $g_4:S^6 \to SU(3)$ be 
its characteristic map. If $g_4$ were a homotopy lifting of $S^3_rh$, 
then there should exist a lifting $g^\prime_4$ by 
$\pi_3:SU(3) \to S^5$ of $S_r^3h :S^6 \to S^5$ {\it i.e.} a 
commuting diagram 
$$\matrix{ & & SU(3) & & \cr & {}^{g^\prime_4} \nearrow & & \searrow 
{}^{\pi_3} & \cr S^6& & \aderecha{S^3_rh}\ & & S^5 \cr}$$ 
with $g_4^\prime \sim g_4$. We have the

{\it Proposition.} $\pi_3$ does {\it not} lift $S^3_rh$.

{\it Proof.} We will show that the homomorphism $\pi_{3*}:\Pi_6(SU(3))\to
\Pi_6(S^5)$ is zero. This implies that any map $S^6 
\aderecha{f}\ S^5$ which factorizes through $\pi_3$ {\it i.e.} a map   
for which there exists a map $S^6 \aderecha{g}\ SU(3)$ such that $\pi_3 
\circ g\sim f$, is null-homotopic. The result now follows from this since 
$S^3_rh$ is essential. 

Consider the long exact homotopy sequence (Steenrod, 1951) of the principal 
bundle $SU(2)\to SU(3) \aderecha{\pi_3}\ S^5$:
$$...\to \Pi_6(SU(3))\aderecha{\pi_{3*}}\ \Pi_6(S^5) \aderecha{\delta}\ 
\Pi_5(S^3)\aderecha{\iota_*}\ \Pi_5(SU(3)) \to ...$$
This gives an exact sequence:
$$...\to \Z_6 \aderecha{\beta}\ \Z_2 \aderecha{\gamma}\ \Z_2 
\aderecha{\alpha}\ \Z \to...$$ where we called $\beta$, $\gamma$ and 
$\alpha$ the homomorphisms corresponding to $\pi_{3*}$, $\delta$ and 
$\iota_*$, respectively. Since $\Z_2$ is a torsion group and $\Z$ is 
torsion free, then the homomorphism $\alpha $ is zero. Therefore 
$\gamma$  is an isomorphism 
{\it i.e.} $ker(\gamma )=ker(\delta )=\{0\}=Im(\pi_
{3*})$ {\it i.e.} $\pi_{3*}=0.$   QED

\subsection{4.2. $(H,f)$-structures}

Let $H$ and $G$ be topological groups (e.g. Lie groups), $\xi _H:H \to 
PH \aderecha{\pi _H} BH$ and $\xi _G:G \to PG \aderecha{\pi _G} BG$ their 
universal bundles, and $H \aderecha{f} G$ a topological group 
homomorphism. Then the action $\bar f:H\times G \to G$, $\bar f(h,g) =f(h)g$ 
induces the associated principal $G$-bundle $(\xi _H)_G: G \to PH \times _H 
G \to BH$ with total space $PH \times _H G= \{[a,g]\}_
{(p,g)\in PH \times G}$, $[a,g]=\{(ah,f(h^{-1})g)\}_{h\in H}$, 
action $(PH \times _H G)\times G \to PH\times _H G$ given by $[a,g]\cdot 
g^\prime =[a,gg^\prime ]$, and projection $(\pi_H)_G([a,g])=\pi_H(a)$.  
$PH\times _H G$ is isomorphic to the pull-back 
bundle $(Bf)^*(PG)$, where the induced function $Bf:BH \to BG$ is uniquely 
defined up to homotopy. 

If $HTop$ is the category of paracompact topological spaces and homotopy 
classes of maps, and $Set$ is the category of sets and functions, then  
for each topological group $K$ there are two cofunctors $k_K$ and $[$  $,BK]$ 
from $HTop$ to $Set$ such that, for each topological group 
homomorphism $H \aderecha{f} G$ there are natural transformations $f_*:
k_H \to k_G$ and $Bf_*:[$  $,BH] \to [$  $,BG]$, and natural equivalences 
$\psi _H$ and $\psi _G$ which make the following functorial diagram 
commutative: $$\matrix{k_H & \aderecha{f_*} & k_G\cr \psi _H \uparrow & & 
\uparrow \psi _G \cr [$  $,BH] & \aderecha{Bf_*} & [$  $,BG] \cr }$$
So, for each topological space $X$ the following set theoretic diagram 
commutes: $$\matrix{k_H(X) & \aderecha{f_*} & k_G(X) \cr \psi_H \uparrow \cong
& & \cong \uparrow \psi_G \cr [X,BH] & \aderecha{Bf_*} & [X,BG] \cr }$$
where: $k_K(X)$= $\{$isomorphism classes of $K$-bundles over $X \}$, $
[X,BK]$= $\{$homotopy classes of maps from $X$ to $BK \}$, $\psi _K (
[\alpha ])=[\alpha ^*(PK)]$, $f_*([\eta ])=[\xi]$ with $\eta :H \to E 
\aderecha{q} X$ and $\xi :G \to E \times _H G \aderecha{\bar q} X$, and
$Bf_*([\alpha ])=[Bf \circ \alpha]$. If $[\xi ] \in k_G(X)$, then 
$f_*^{-1}(\{[\xi ] \})$ is the set of $(H,f)$-{\it structures} on $\xi $; 
this set can be empty. So, $\xi $ has a $(H,f)$-structure if and only if 
there exists a map $\alpha :X \to BH$ such that $Bf \circ \alpha \sim F$, 
where $F$ is the classifying map of $\xi $. One can show that this 
definition is equivalent to the existence of a $G$-bundle isomorphism 
$$\matrix{
E \times _H G   & \aderecha{\bar {\varphi }} & P \cr
\bar q \searrow &                            & \swarrow \pi \cr
                & X                          & \cr }$$
where $H \to E \aderecha{q} X$ is a
principal $H$-bundle or, equivalently,  
to the bundle map
$$\matrix{
E \times H        & \aderecha{\varphi \times f} & P \times G \cr
\kappa \downarrow &                             & \downarrow \psi \cr
E                 & \aderecha{\varphi}          & P \cr
q \searrow        &                             & \swarrow \pi \cr
                  & X                           &  \cr}$$
where $\varphi
=\bar {\varphi }\circ \varphi _f$ with $\varphi _f:E \to E\times _H G$ given by $
\varphi _f(a)=[a,e]$ ($e$ is the unit of $G$) (Aguilar and Socolovsky, 1997a). One  
says that $(E,\varphi)$ {\it is a} $(H,f)$-{\it structure on} $G \to P 
\aderecha{\pi } X$.

In the case of smooth bundles, if the Lie group homomorphism $H \aderecha{f} G$ 
is an {\it embedding} {\it i.e.} an 
injective immersion, then $E$ is called a {\it reduction} of $P$ to $H$. 
In this setting, one has the following

{\it Proposition.} If $f$ is an embedding, then $\varphi $ is also an 
embedding. 

{\it Proof.} Since $\varphi =\bar \varphi \circ \varphi _f$ and $ \bar 
\varphi $ is a diffeomorphism, then $\varphi $ is an embedding if and only 
if $\varphi _f$ is an embedding; we shall show that $\varphi_f$ is an  
embedding. i) $\varphi _f$ 
is injective: Let $\varphi _f (a_1)=\varphi _f (a_2)$ {\it i.e.} $[a_1,e]=
[a_2,e]$, since $[a,e]=\{(ah,f(h^{-1}))\}_{h \in H}$ then there must 
exist $h \in H$ such that $(a_1,e)=(a_2 h,f(h^{-1})$ {\it i.e.} 
$a_1=a_2 h$ and $f(h^{-1})=e$, but $f$ is injective, so $h^{-1}=h=e^\prime$,  
the identity in $H$, and then $a_1=a_2$. ii) 
$d\varphi _f$ is injective at each $a_0 \in E$:  
Consider the commutative diagram
$$\matrix{
  &        & E\times G             &             \cr
  &i\nearrow &                       & \searrow p   \cr
E &        & \aderecha{\varphi _f} & E\times _H G \cr}$$
where $i(a)=(a,e)$ and $p(a,e)=[a,e]$. One can prove that $H \to E\times G 
\aderecha{p} E\times _H G$ is a principal $H$-bundle, so fixing $(a_0,e) 
\in E\times G$ there is a map $\alpha _{(a_0,e)}:H \to E\times 
G$, given by $\alpha _{(a_0,e)}(h)=(a_0,e)\cdot h=(a_0h,f(h^{-1}))$, in 
particular 
$\alpha _{(a_0,e)}(e^\prime)=(a_0,e)$. One then has the following diagram 
of vector spaces:
$$\matrix{0 \to 
T_{e^\prime }H \aderecha{(d\alpha _{(a_0,e)})_{e^\prime}}
 & T_{(a_0,e)}(E\times G) \aderecha{(dp)_{(a_0,e)}}
 & T_{[a_0,e]} (E\times _H G) \to 0 \cr\noalign{\vskip3pt}
& (di)_{a_0} \uparrow
 & \nearrow (d\varphi _f )_{a_0} &  \cr\noalign{\vskip3pt}
& T_{a_0}E   & &\cr}$$
where the horizontal sequence is exact and the triangle commutes. 
If $p_1$ and $p_2$ are respectively the 
projections of $E\times G$ onto $E$ and $G$, then $(di)_{a_0}(v)=
((d(p_1 \circ i))_{a_0} (v),(d(p_2 \circ i))_{a_0} (v))=(v,0)$ since 
$p_1 \circ i (a)=a$ {\it i.e.} $p_1 \circ i=id_E$ and $p_2 \circ i(a)=p_2
(a,e)=e$ {\it i.e.} $p_2 \circ i=const.$, and therefore $(d(p_1 \circ i))
_{a0}=(d(id_E ))_{a_0}=id_{T_{a_0}E}$ and $(d(p_2 \circ i))_{a_0}=
(d(const.))_{a_0}=0$. On the other hand, $(d\alpha _{(a_0,e)})_{e^\prime}
(w)=((d(p_1 \circ \alpha _{(a_0,e)}))_{e^\prime}(w),(d(p_2 \circ \alpha _
{(a_0,e)}))_{e^\prime}(w))=((d\alpha_{a_0})_{e^\prime}(w),(df)_{e^\prime}
\circ (d\gamma)_{e^\prime}(w))$ where we have used $p_1 \circ \alpha 
_{(a_0,e)}(h)=p_1(a_0 h,f(h^{-1}))=a_0 h:=\alpha _{a_0}(h)$, $p_2 \circ 
\alpha _{(a_0,e)}(h)=p_2(a_0 h,f(h^{-1}))=f\circ \gamma (h)$ with $\gamma 
:H \to H$ given by $\gamma (h)=h^{-1}$. Let $(r,s) \in Im((di)_{a_0})
\cap Im((d\alpha _{(a_0,e)})_{e^\prime}))$, then $s=0$ and so $0=(df)_{
e^\prime}((d\gamma)_{e^\prime}(w))$, therefore $(d\gamma)_{e^\prime}(w)=0$ 
because $f$ is an immersion and, since $\gamma $ is a diffeomorphism, 
$w=0$, so $r=(d\alpha_{a_0})_{e^\prime}(0)=0$ {\it i.e.} $Im((di)_{a_0})
\cap Im((d\alpha_{(a_0,e)})_{e^\prime})=\{0\}$. Finally, let $v \in  
ker ((d\varphi _f)_{a_0})$, then $0=(dp)_{(a_0,e)}((di)_{a_0})(v))$ {\it 
i.e.} $(di)_{a_0}(v)\in ker((dp)_{(a_0,e)})=Im((d\alpha_{(a_0,e)})_
{e^\prime})$ {\it i.e.} $(di)_{a_0}(v)=0$. Since $i$ is an embedding, 
then $v=0$ {\it i.e.} $(d\varphi _f)_{a_0}$ is one-to-one.   QED

{\it Remark.} One often finds in the literature (Kobayashi and Nomizu, 1963; 
Trautman, 1984) that to define a reduction to a Lie subgroup $H\subset G$, 
$\varphi$ is required to be an embedding. The proposition above shows 
that this is a consequence of the fact that $H\to G$ is an embedding.

\subsection{4.3. $SU(n)\to SU(n+1)\aderecha{\pi_{n+1}} S^{2n+1}$ as an 
$(SU(n),\iota)$-structure on $U(n) \to U(n+1) \aderecha{p_{n+1}} S^{2n+1}$}

{\it Proposition.} For $n=1,2,3,...$, the bundle $\pi_{n+1}$  is a reduction of 
the bundle $p_{n+1}$ {\it i.e.} one has the $U(n)$-bundle isomorphism given by the 
commutative diagram 
$$\matrix{
(SU(n+1)\times _{SU(n)} U(n))\times U(n) & \aderecha{\bar {\varphi }\times id} & U(n+1)\times U(n) \cr
\lambda \downarrow                       &                                   & \downarrow \psi \cr
SU(n+1)\times _{SU(n)} U(n)              & \aderecha{\bar {\varphi}}           & U(n+1)  \cr
q_{n+1} \searrow                         &                                   & \swarrow p_{n+1} \cr
                                         & S^{2n+1}                          & \cr}$$
(The actions $\lambda$ and $\psi$, the diffeomorphism $\bar {\varphi}$, and 
the projections $q_{n+1}$ and $p_{n+1}$ are given below.)

{\it Proof.} 

Consider the inclusion $SU(n+1) \aderecha{\varphi} U(n+1)$; one can easily 
show that this is a bundle map between the principal $SU(n)$-bundle 
$\pi_{n+1}$ and the principal $U(n)$-bundle $p_{n+1}$. From the general theory 
mentioned above, $\bar \varphi$ is a bundle isomorphism given by 
$\bar \varphi ([D,A])=Dj(A)$ where $j$ is the inclusion $U(n)\to U(n+1)$ 
given by $j(A)=
\pmatrix{A & 0 \cr 0 & 1 \cr}$. The inverse of $\bar \varphi$ is given 
as follows: if $C \in 
U(n+1)$ then $C=Dl(det C)$ with $D=C(l(det C))^{-1} \in SU(n+1)$ and $
l:U(1) \to U(n+1)$ is the inclusion $l(z)=\pmatrix{z & 0 \cr 0 & I \cr}$, 
then $[D,l(det C)]= \bar \varphi ^{-1}(C)$.   QED 

\subsection{4.4. Proof of the main result}

{\it Proposition.} Let $H \aderecha{f} G$ be a homomorphism of path 
connected topological groups. Then, for each $n=1,2,3,...$, the following 
diagram of sets commutes:
$$\matrix{
[S^{2n},G] & \aderecha{\mu _{G \#}} & [S^{2n},\Omega BG] & \aderecha{adj_{G*}} & 
[S_rS^{2n},BG] & \aderecha{\psi _G} & k_G(S^{2n+1}) \cr 
\uparrow f_{\#} & & \uparrow \Omega Bf_{\#} & & \uparrow Bf_* & & \uparrow f_* \cr 
[S^{2n},H] & \aderecha{\mu _{H \#}} & [S^{2n},\Omega BH] & \aderecha{adj_{H*}} &
[S_rS^{2n},BH] & \aderecha{\psi _H} & k_H(S^{2n+1}) \cr}$$

$k_K$, $[$  $,$  $]$, $f_*$, $Bf_*$ and $\psi _K$ have been defined before, 
$\Omega BK$ is the loop space of $BK$, and $f_{\#}$, $\mu _{K \#}$, $\Omega Bf_{\#}$ 
and $adj_{K*}$ are given by $f_{\#}([\delta ])=[f \circ \delta ]$, $\mu _{K \#}
([\sigma ])=[\mu _K \circ \sigma ]$ ($\mu _K$ is defined below), 
$\Omega Bf_{\#}([\kappa ])=[\Omega Bf 
\circ \kappa]$ where $\Omega Bf:\Omega BH \to \Omega BG$ is given by 
$\Omega Bf(\gamma )=Bf \circ \gamma$, and $adj_{K*}([\alpha ])=[adj_K(
\alpha )]$ with $adj_K(\alpha )([z,t])=\alpha (z)(t)$, $t \in [0,1]$. 
$[S^{2n},K]$ are the clutching maps for the $K$-principal bundles over 
$S^{2n+1}$. 

{\it Proof.} The commutativity of the right square has been proved in 
Section 4.2, with $S^{2n+1}=X$. The natural equivalence $adj_K$ is given 
by the {\it exponential law} in function spaces (Aguilar, Gitler, and 
Prieto, 1998). Finally, there exist {\it weak homotopy equivalences} $\mu _K$ 
such that the diagram 
$$\matrix{
H               & \aderecha{f}         & G \cr
\mu _H \uparrow &                      & \uparrow \mu _G \cr
\Omega BH       & \aderecha{\Omega Bf} & \Omega BG \cr}$$
commutes up to homotopy (Switzer, 1975).   QED

{\it Proposition.} For even $n$, $n \geq 2$, the clutching map $g_{n+1}$ of 
the principal bundle $SU(n) \to SU(n+1) \aderecha{\pi _{n+1}} S^{2n+1}$ is 
a homotopy lifting of the $(2n-3)-th$ reduced suspension of the Hopf map $h$. 
For odd $n$, $n \geq 3$, $\pi_n \circ g_{n+1}$ is inessential.

{\it Proof.} We apply the previous proposition to the case $H=SU(n)$, 
$G=U(n)$, and $f=\iota$ (the inclusion), for $n=2,3,...$. By the proposition  
in Section 4.3, $[\xi ]=\iota _*([\eta ])$ with $\xi :U(n) \to U(n+1)  
\aderecha{p_{n+1}} S^{2n+1}$ and $\eta :SU(n) \to SU(n+1) \aderecha{\pi 
_{n+1}} S^{2n+1}$. Then one has the commutative diagram 
$$\matrix{
[T^\prime _{n+1}] & \in & [S^{2n},U(n)]        & \aderecha{\mu } &k_{U(n)}(S^{2n+1})   & \ni & [\xi ] \cr
                  &     & \iota _{\#} \uparrow &                 &\iota _* \uparrow    &     &        \cr
[g_{n+1}]         & \in & [S^{2n},SU(n)]       & \aderecha{\nu } & k_{SU(n)}(S^{2n+1}) & \ni & [\eta ] \cr}$$
where $\mu =\psi _{U(n)} \circ adj_{U(n)*} \circ \mu _{U(n) \#}$ and 
$\nu =\psi _{SU(n)} \circ adj_{SU(n)*} \circ \mu _{SU(n) \#}$ are bijections, 
and $T^\prime _{n+1}$ is the clutching map for the principal bundle $\xi $ 
(Steenrod, 1951). Then $[T^\prime _{n+1}]=\mu ^{-1}([\xi ])=\mu ^{-1}(
\iota _*([\eta ]))=\mu ^{-1} \circ \iota _* (\nu ([g_{n+1}]))=
\mu ^{-1} \circ \iota _* \circ \nu ([g_{n+1}])=\iota _{ \#}([g_{n+1}])=
[\iota \circ g_{n+1}]$ and therefore $T^\prime _{n+1} \sim \iota \circ 
g_{n+1}$. 

Consider the diagram (not necessarily commutative) 
$$\matrix{
       &                          & U(n+1)                &              & \cr
       &                          & \downarrow            &              & \cr
       &                          & U(n)                  &              & \cr
       & T^\prime _{n+1} \nearrow &                       & \searrow p_n & \cr
S^{2n} &                          &\aderecha{S_r^{2n-3}h} &              &S^{2n-1} \cr}$$
Steenrod (Steenrod, 1951) proved that for $n$ even, $n \geq 2$, $p_n 
\circ T^\prime _{n+1} \sim S_r^{2n-3}h$ {\it i.e.} the diagram commutes 
up to homotopy, while for $n$ odd, $n \geq 3$, $p_n \circ T^\prime _
{n+1} \sim const.$ Then, $p_n \circ T^\prime _{n+1} \sim p_n \circ 
(\iota \circ g_{n+1})=(p_n \circ \iota )\circ g_{n+1}=\pi _n \circ 
g_{n+1}$
$$
\sim
\bigg\{ \matrix{S_r^{2n-3}h, \ n \ even \cr const., \ n \ odd \cr}
$$
QED

\section{5. $S^2$ and relativity}

As is well known, the Lorentz group, the group of linear 
transformations of Minkowski space-time which preserves the scalar 
product $<x,y>=x^T \eta y$ where $\pmatrix{1 & 0 & 0 & 0 \cr 
0 & -1 & 0 & 0 \cr 
0 & 0 & -1 & 0 \cr 0 & 0 & 0 & -1 \cr}$ is the Minkowskian metric, 
is a subgroup of 
the symmetry group of several gauge theories of gravity (Hehl {\it et 
al}, 1976; Basombr\'\i o, 1980). This means that $O(3,1)$ is a 
subgroup of the structure group of the corresponding principal bundles. 
The relationship between these theories and the 2-sphere (the Riemann 
sphere $\C \cup \{ \infty \}$) comes from the fact that there is a 
canonical isomorphism between the connected component of $O(3,1)$, the 
proper orthochronous Lorentz group $SO^0(3,1)$ and the group of 
conformal (Moebius) transformations of $S^2$, $Conf(S^2)$. We recall 
that $Conf(S^2)$ is the set of all invertible transformations of the 
Riemann sphere which preserves the angles between curves and that at 
each point multiply all the tangent vectors by a fixed positive number. 

Let $g=\pmatrix{a & b \cr c & d \cr}$ be an element of $GL_2(\C)$, we 
define a Moebius transformation $m:S^2 \to S^2$ as follows: if $c \neq 
0$, then 
$$
z \mapsto
\cases{{az+b\over cz+d} & if $z \neq -d/c$\cr
\infty & if $z= -d/c$\cr}
$$
and  
$$\infty \mapsto a/c;$$

and, if $c=0$, then 
$$
\left\{\matrix{z\mapsto {a\over d}z + {b\over d}\cr
\infty \mapsto \infty\cr}\right.
$$

It is then easy to verify that the following diagram commutes: 
$$\matrix{
          &               & \Z _2           &                  & \cr
          &               & \downarrow      &                  & \cr
          &               & SL_2(\C)        &                  & \cr
          & \psi \swarrow &                 & \searrow \lambda & \cr
SO^0(3,1) &               &\aderecha {\tau} &                  & Conf(S^2) \cr}$$
where: i) the projections $\psi $ and $\lambda $ are two-to-one 
group homomorphisms, respectively given by $\psi \pmatrix{a & b \cr 
c & d \cr}$ $$=\pmatrix{{\vert a \vert ^2 + \vert b \vert  
^2 + \vert c \vert ^2 + \vert d \vert ^2 \over 2} & Re(a \bar b+ c \bar d) & 
Im(a \bar b + c \bar d) & {\vert a \vert ^2 - \vert b \vert 
^2 + \vert c \vert ^2 - \vert d \vert ^2 \over 2} \cr Re(a \bar c + b \bar d) & 
Re(a \bar d + b \bar c) & Im(a \bar d - b \bar c) & Re(a \bar c- b 
\bar d) \cr -Im(a \bar c +b \bar d) & Im(a \bar d - b \bar c) & 
Re(a \bar d - b \bar c) & 
-Im(a \bar c - b \bar d) \cr {\vert a \vert ^2 + \vert b 
\vert ^2 - \vert c \vert ^2 - \vert d \vert ^2 \over 2} & Re(a \bar b -c \bar d) 
& Im(a \bar b - c \bar d) & {\vert a \vert ^2 - \vert b \vert  
^2 - \vert c \vert ^2 + \vert d \vert ^2 \over 2} \cr}$$
with $\psi \pmatrix{a & b \cr c & d \cr}\!=\psi \pmatrix{-a & -b \cr 
-c & -d \cr}\!=l$ (Penrose and Rindler, 1984), 
and $\lambda (g/ \sqrt {det g})
\!=m$ with $\lambda (g/ \sqrt {det g})= \lambda (-g/ \sqrt {det g})$; and 
ii) $\tau (l) = m$ is the desired isomorphism. $SL_2(\C) \aderecha{\psi } 
SO^0(3,1)$ and $SL_2(\C) \aderecha{\lambda } Conf(S^2)$ are $\Z_2$- principal 
bundles. 

Thus we conclude that the symmetry group of the standard model $G^\prime _
{SM}$, when gravitation is included, locally contains, as a space, $S^1 \times 
(S^3)^2 \times S^5 \times Conf(S^2)$. 

{\it Remark.} In the framework of the theory of categories, functors, and 
natural 
transformations, some of the geometrical objects of the previous 
sections, e.g. spheres and the Hopf map, have a natural origin. 
This suggests a possible relation between symmetries in nature, and 
therefore conservation laws, and some of the most general mathematical 
concepts. The basic idea is that of a representable functor (Aguilar 
and Socolovsky, 1997).

\section{REFERENCES}

\noindent Aguilar, M., Gitler, S., and Prieto, C. (1998). {\it
Topolog\'\i a Algebraica, Un enfoque homot\'opico}, McGraw Hill,
M\'exico.

\noindent Aguilar, M. A., and Socolovsky, M. (1997a). Reductions and
extensions in bundles and homotopy, {\it Advances in Applied Clifford
Algebras}, {\bf 7 (S)}, 487-494.

\noindent Aguilar, M. A., and Socolovsky, M. (1997b). Naturalness of
the space of States in Quantum Mechanics, {\it International Journal of
Theoretical Physics}, {\bf 36}, 883-921.

\noindent Ashtekar, A., and Schilling, T.A. (1995). Geometry of Quantum 
Mechanics, {\it AIP Conference Proceedings}, {\bf 342}, 471-478.

\noindent Basombr\'\i o, F. G. (1980).A Comparative Review of Certain